\documentclass[twocolumn,epjc3]{svjour3}
\smartqed  
\RequirePackage{graphicx}
\usepackage{widetext}
\usepackage{amsmath,amssymb,amsfonts}
\usepackage{graphicx,graphics}
\usepackage{graphics}
\usepackage{url}
\usepackage{slashed}
\usepackage{color}
\usepackage{mathtools}
\usepackage{nccmath}
\usepackage{ulem}

\graphicspath{{fig/}}
\definecolor{theblue}{RGB}{0,50,230}

\newcommand{\trento}{{\tt{TRENTo}}}

\newcommand{\T}{\tilde{T}}

\newcommand {\avg}[1]{\ensuremath{\langle\kern-1.0pt\langle#1\rangle\kern-1.0pt\rangle}}

\newcommand{\dphi}    {\ensuremath{\Delta\phi}}

\newcommand{\La}{\langle}
\newcommand{\Ra}{\rangle}

\def\bra{\langle}
\def\ket{\rangle}

\newlength\cmsFigWidth

\begin{document}

\title{Collective flow in 2.76 A TeV and 5.02 A TeV Pb+Pb collisions}

\author{Wenbin~Zhao\thanksref{addr1}
        \and
        Hao-jie~Xu\thanksref{addr1}
        \and
        Huichao~Song\thanksref{addr1,addr2,addr3,e1} }

\institute{Department of Physics and State Key Laboratory of Nuclear Physics and Technology, Peking University, Beijing 100871, China\label{addr1}
                  \and
 Collaborative Innovation Center of Quantum Matter, Beijing 100871, China\label{addr2}
               \and
 Center for High Energy Physics, Peking University, Beijing 100871, China\label{addr3}}

\thankstext{e1}{e-mail: Huichaosong@pku.edu.cn}
\date{\today}
\maketitle
\begin{abstract}
In this paper, we study and predict flow observables in 2.76 A TeV and 5.02 A TeV Pb +Pb collisions, using the {\tt{iEBE-VISHNU}} hybrid model with \trento\ and {\tt AMPT} initial conditions and with different forms of the QGP transport coefficients. With properly chosen and tuned parameter sets, our model calculations can nicely describe various flow observables in 2.76 A TeV Pb +Pb collisions, as well as the measured flow harmonics of all charged hadrons in 5.02 A TeV Pb +Pb collisions. We also predict other flow observables, including $v_n(p_T)$ of identified particles, event-by-event $v_n$ distributions, event-plane correlations, (Normalized) Symmetric Cumulants, non-linear response coefficients and $p_T$-dependent factorization ratios, in 5.02 A TeV Pb+Pb collisions.  We find many of these observables remain approximately the same values as the ones in 2.76 A TeV Pb+Pb collisions.  Our theoretical studies and predictions could shed light to the experimental investigations in the near future.
\end{abstract}

\section{Introduction}

\quad \ At extreme high temperature and density, the nuclear matter can experience a phase transition and form the quark-gluon plasma (QGP). The main goals of the relativistic heavy-ion collisions at Relativistic Heavy Ion Collider (RHIC) and the  Large Hadron Collider (LHC) are to create the QGP and to explore its properties~\cite{Rev-Arsene:2004fa,Gyulassy:2004vg,Muller:2006ee}. Since the running of RHIC in 2000, strong evidences have been gradually accumulated for the creation of the QGP, including jet quenching, strong collective flow and the valance quark scaling of the elliptic flow~\cite{Rev-Arsene:2004fa,Gyulassy:2004vg,Muller:2006ee}. Hydrodynamics and hybrid models are successful tools to simulate the collective expansion of the QGP fireball and to study various flow observable at RHIC and the LHC~\cite{Teaney:2009qa,Romatschke:2009im,Huovinen:2013wma,Heinz:2013th,Gale:2013da,Song:2017wtw}. The past research has revealed that the created QGP fireballs fluctuate event-by-event and behave like nearly perfect liquids with very small specific shear viscosity~\cite{Heinz:2013th,Gale:2013da,Song:2017wtw,Song:2013gia,Luzum:2013yya,Jia:2014jca}.

In the past few years, various flow observables have been extensively measured and studied in 2.76 A TeV Pb+Pb collisions, including the integrated and differential flow harmonics~\cite{ALICE:2011ab,ATLAS:2012at,Alver:2010dn,Song:2011qa,Song:2013qma,Xu:2016hmp,Gale:2012rq}, the event-by-event $v_n$ distributions~\cite{Gale:2012rq,Aad:2013xma,Yan:2014afa,Zhou:2015eya}, the event-plane correlations~\cite{Aad:2014fla,Qiu:2012uy,Bhalerao:2013ina,Teaney:2013dta,Bhalerao:2014xra}, and the correlations between different flow harmonics (Symmetric Cumulants)~\cite{Bhalerao:2014xra,Aad:2015lwa,ALICE:2016kpq,Niemi:2015qia,Giacalone:2016afq,Zhu:2016puf,Qian:2016pau},
the $p_T$ or $\eta$-dependent de-correlations of the flow vector~\cite{Heinz:2013bua,Gardim:2012im,Khachatryan:2015oea,Zhou:2014bba,Pang:2014pxa,Pang:2015zrq,Xiao:2015dma,Ma:2016fve} and etc.. Many of these flow observables reflect the information on the event-by-event initial state fluctuations and the non-linear evolution of the system, which provide constraints for the initial condition models and the QGP transport coefficients.  For example, it was found that the event-by-even $v_n$ distributions  mostly follow the the event-by-even $\varepsilon_n$ distributions of the initial state for n=2 and 3, which does not favor the traditional {\tt MC-Glauber} and {\tt MC-KLN} models with nucleon position fluctuations~\cite{Gale:2012rq,Aad:2013xma}. Based on eikonal entropy deposition via a reduced thickness function, Moreland and his collaborators constructed a parametric \trento\ model that could match various initial conditions with tunable parameters~\cite{Moreland:2014oya}.  Using  \trento\ initial conditions, the Duke and OSU group has performed massive data simulations of {\tt iEBE-VISHNU} hybrid model and systematically evaluated the measured multiplicity, mean $p_T$ and integrated $v_n$ in 2.76 A TeV Pb+Pb collisions. The extracted temperature dependent specific shear viscosity $\eta/s(T)$ is an approximately linear function with a minimum value close to the KSS bound  near $T_c$~\cite{Bernhard:2016tnd}.
The early hydrodynamic or hybrid model simulations, using either {\tt IP-Glasma}~\cite{Gale:2012rq}, or {\tt AMPT}~\cite{Xu:2016hmp} or {\tt EKRT} initial conditions~\cite{Niemi:2015qia}, can also nicely fit the integrated and differential flow harmonics with a constant or temperature dependent $\eta/s$, close to the KSS bound near $T_c$.
In fact, the flow harmonics $v_n$ are not sensitive to the details of the initial condition models as along as the balanced initial eccentricities can be produced with some tunable parameters.  Other flow measurements, e.g., the event-plane correlations, Symmetric Cumulants, non-linear response coefficients, the de-correlation of the flow vector and etc., could reveal more details on initial state fluctuations and the non-linear hydrodynamic response~\cite{Aad:2014fla,Qiu:2012uy,Bhalerao:2013ina,Teaney:2013dta,Bhalerao:2014xra,Aad:2015lwa,ALICE:2016kpq,Niemi:2015qia,Giacalone:2016afq,Zhu:2016puf,Qian:2016pau,Heinz:2013bua,Gardim:2012im,Khachatryan:2015oea,Zhou:2014bba,Pang:2014pxa,Pang:2015zrq,Xiao:2015dma,Ma:2016fve}. A systematic study of these flow observables will help us to test the model calculations and the extracted QGP viscosity as well as to further evaluate and constrain the initial condition models.

Recently, the ALICE collaboration has measured the integrated and differential flow harmonics of all charged hadrons in 5.02 A TeV Pb +Pb collisions~\cite{Adam:2016izf}. It was found, with the collision energies raised from 2.76 A TeV to 5.02 A TeV, $v_2$,  $v_3$ and $v_4$ slightly increase with the increase of average transverse momentum, as predicted by early hydrodynamic calculations~\cite{Niemi:2015voa,Noronha-Hostler:2015uye}. In this paper, we will implement {\tt{iEBE-VIHSNU}}
hybrid model with  \trento\ and {\tt{AMPT}} initial conditions to study and predict various flow observable in 2.76 A TeV and 5.02 A TeV Pb+Pb. Instead of predicting the flow harmonics $v_n$ of all charged hadrons at 5.02 A TeV (which has been done in~\cite{Niemi:2015voa,Noronha-Hostler:2015uye}), we use these available data to fix the free parameters in the {\tt{iEBE-VIHSNU}} simulations and then make predictions for other flow observables, including the differential flow harmonics $v_n(p_T)$ of identified hadrons, the event-by-event $v_n$ distributions, the event-plane correlations, the Symmetric Cumulants, non-linear response coefficients and the $p_T$-dependent factorization ratios.
We have noticed that the MC-grill group also predicted various flow observables in 5.02 A TeV Pb+Pb collisions, using MUSIC simulations with the IP-Glasma initial conditions~\cite{McDonald:2016vlt}. Compared with their calculations~\cite{McDonald:2016vlt} and other early investigations~\cite{Niemi:2015voa,Noronha-Hostler:2015uye}, our predictions are more complete, which are also on time and can be measured in the near future. For example, the Symmetric cumulants and non-linear response coefficients in 5.02 A TeV Pb+Pb collisions are firstly predicted in this paper, which have not been done elsewhere as far as we know. Secondly, the parameters in {\tt{iEBE-VIHSNU}} are fine tuned to fit the published soft hadron data, which give more reliable predictions for these un-measured flow observables. For example, our descriptions of $v_n(p_T)$ of all charged hadrons are better than the ones in~\cite{McDonald:2016vlt}. Correspondingly, the predicted flow harmonics of identified hadrons are also more reliable. Besides, it is worthwhile to investigate the same flow observables using the hydrodynamic calculations with different initial conditions, which could help us to understand the details of the initial state fluctuations and may help us to locate some certain flow observables to further constrains the initial conditions.

  This paper is organized as the following: Sec.~2 introduces
the {\tt iEBE-VISNU} hybrid model and the set-ups of calculations with \trento\ and {\tt AMPT} initial conditions. Sec.~3 introduces the methodology to calculate various flow observables.  Sec.~4 presents and discusses the calculated and predicted flow observables in 2.76 A TeV and 5.02 A TeV  Pb+Pb collisions. Sec.~5 summarizes and concludes.

\section{The model and set-ups of the calculations}
\subsection{\textbf{\textbf{\tt iEBE-VISHNU} hybrid model}}

\quad \ In this paper, we will implement {\tt iEBE-VISHNU} hybrid model to study and predict various flow observables in 2.76 A and 5.02 A TeV Pb+Pb collisions.  {\tt iEBE-}
{\tt VISHNU}~\cite{Shen:2014vra} is an event-by-event version of the  {\tt VISHNU} hybrid model~\cite{Song:2010aq}, which combines (2+1)-d viscous hydrodynamics {\tt VISH2+1}~\cite{Song:2007fn,Song:2007ux,Song:2009gc} to describe the expansion of the QGP fireball with a hadron cascade model ({\tt UrQMD})~\cite{Bleicher:1999xi,Bass:1998ca} to simulate the succeeding evolution of the hadron resonance gas.

In the hydrodynamics part, {\tt iEBE-VISHNU} solves the transport equations for energy-momentum tensor $T^{\mu \nu}$ and the 2nd order Israel-Stewart equations for shear stress tensor $\pi^{\mu \nu}$ and bulk pressure $\Pi$~\cite{Song:2007fn,Song:2007ux,Song:2009gc}:
%
\begin{eqnarray}
&& \partial_\mu T^{\mu \nu}(x)=0,  \ \ \   \
 T^{\mu \nu}=e u^{\mu}u^{\nu}-(p +\Pi)\Delta^{\mu\nu}+\pi^{\mu \nu}, \nonumber \\
&&\dot{\Pi}=-\frac{1}{\tau_{\Pi}}\bigg[\Pi+\zeta \theta+\Pi \zeta T
\partial_{\mu} \big( \frac{\tau_\Pi u^{\mu}}{2\zeta T} \big) \bigg],\\
&&\Delta ^{\mu \alpha} \Delta ^{\nu \beta}\dot{ \pi}_{\alpha \beta}
=-\frac{1}{\tau_{\pi}}\bigg[\pi^{\mu\nu}-2\eta \nabla ^{\La \mu}u
^{\nu\Ra}
+ \pi^{\mu\nu} \eta T
\partial_{\alpha} \big( \frac{\tau_\pi u^{\alpha}}{2 \eta T} \big)
\bigg],   \nonumber
\end{eqnarray}
%
where $e$, $p$ and $T$ are the local energy density, pressure and temperature,
and $u^{\mu}$ is the flow 4-velocity. $\Delta^{\mu\nu} =g^{\mu\nu}{-}u^\mu u^\nu$,
$\nabla ^{\La \mu}u ^{\nu\Ra}=\frac{1}{2}(\nabla^\mu u^\nu+\nabla^\nu
u^\mu)-\frac{1}{3}\Delta^{\mu\nu}\partial_\alpha u^\alpha$ and
$\theta=\partial \cdot u$.  $\eta$ is the shear viscosity, $\zeta$ is the
bulk viscosity and $\tau_{\pi}$, $\tau_{\Pi}$ are the corresponding relaxation times.
Here, we neglect the equations for net charge current and heat flow since we focus on
the soft physics at the LHC, where both net baryon density and heat conductivity
are negligible. With a Bjorken approximation $v_z=z/t$~\cite{Bjorken:1982qr},
the above equations can be written in a 2+1-d form with longitudinal boost invariance~\cite{Song:2007ux,Song:2009gc,Heinz:2005bw},
which largely increase the numerical efficiency when compared with the full 3+1-d simulations.

For the hydrodynamic simulations,  one needs to input an equation of state (EoS), $P=P(e)$, to close the system.  Following~\cite{Bernhard:2016tnd}, we implement a state-of-art EoS that matches the recent lattice EoS at zero baryon
density from the HotQCD collaboration~\cite{Bazavov:2014pvz} and the hadron resonance gas EoS using a smooth
interpolation function.

In the hybrid model, the switch between hydrodynamics and hadron cascade simulations
is realized by a particle event generator, which converts the hydrodynamic outputs on
a switching hyper-surface
into various hadrons with specific momentum and
position for the succeeding {\tt UrQMD} simulations.
More specifically, such Monte Carlo event generator is constructed
according to the differential Cooper-Frye formula~\cite{Song:2010aq}:
\begin{ceqn}
\begin{eqnarray}\label{dis}
 E\frac{d^3N_i}{d^3p}(x) &=& \frac{g_i}{(2\pi)^3}
  p\cdot d^3\sigma(x)\, f_i(x,p),
\end{eqnarray}
\end{ceqn}
where $f_i$ is the distribution function of particle $i$ which includes both
equilibrium and non-equilibrium contributions $f_i=f_{i0}+\delta f_i$.  $d^3\sigma(x)$ is a volume element of the switching hypersurface $\Sigma$, which is generally defined by a constant switching temperature $T_{swith}$. Following~\cite{Bernhard:2016tnd}, $T_{switch}$ is set to 148 MeV and the non-equilibrium distribution function is taken the form $\delta f= \delta f_{shear}=f_0 \bigl(1{\mp}f_0\bigr)\frac{p^\mu p^\nu \pi_{\mu\nu}}{2T^2\left(e{+}p\right)}$~\footnote{Note that the bulk viscous correction $\delta f_{bulk}$ is neglected here. In fact, $\delta f_{bulk}$ has a variety of forms, which more or less influences the flow observables when bulk pressure or transverse momentum become large~\cite{Dusling:2011fd,Noronha-Hostler:2013gga}.  To avoid such uncertainties for the massive data fitting, Ref.~\cite{Bernhard:2016tnd} directly set $\delta f_{bulk}=0$ in the particle event generator of {\tt iEBE-VISHNU}.
For our simulations with \trento\ initial condition, we input the same parameterizations for
specific shear and bulk viscosity  (para-I in Fig.~1) and thus set $\delta f=\delta f_{shear}$ as~\cite{Bernhard:2016tnd}. For the {\tt AMPT} initial condition, we input a constant specific shear viscosity and zero bulk viscosity (para-II in Fig.~1) in the {\tt iEBE-VISHNU} simulations, which does not need the additional $\delta f_{bulk}$ corrections for $\delta f$.}.

After conversing the fluid into various hadrons, the evolution of the hadron matter is simulated  by the Ultra-relativistic Quantum Molecular Dynamics ({\tt  UrQMD}) through solving the Bolzmann equations~\cite{Bleicher:1999xi,Bass:1998ca}:
\begin{ceqn}
\begin{eqnarray}
 \frac{d f_i(x,p)}{dt}= C_i(x,p),
\end{eqnarray}
\end{ceqn}
where $f_i(x,p)$ is the distribution function of hadron species $i$ and $C_i(x,p)$ is the corresponding collision terms.
According to these equations, the produced hadrons propagate along classical trajectories, together with the elastic, inelastic  scatterings and resonance decays. When all the interactions cease, the evolution stops and final information of produced hadrons are output to be further analyzed and compared with the experimental data.

\begin{figure}[t]
  \includegraphics[scale=0.43]{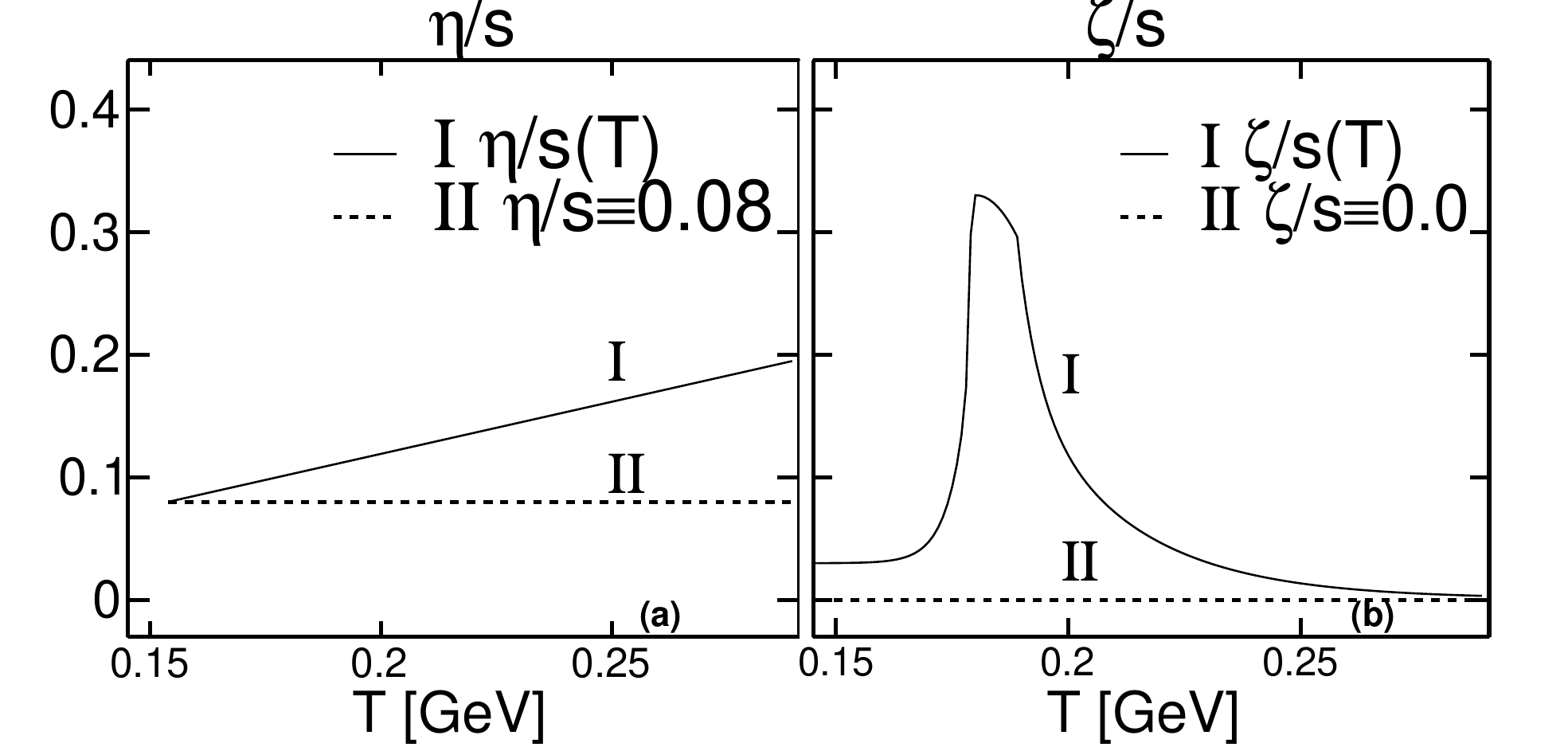}
  \vspace{1mm}
  \caption{Two sets of specific shear viscosity $\eta/s$  and specific bulk viscosity $\zeta/s$ as a function of temperature, used in {\tt iEBE-VISHNU} simulations with \trento\ initial condition (para-I) and {\tt{AMPT}} initial condition (para-II).}
  \label{vis}
\end{figure}

\subsection{\textbf{Set-ups}}
\quad \ In this paper, we will implement two different initial conditions, called \trento\ and {\tt AMPT}, in the {\tt iEBE-VISHNU} simulations. In this sub-section, we will briefly introduce these two initial conditions and the set-ups of related parameters for the simulations in Pb+Pb collisions at 2.76 A TeV and 5.02 A TeV.\\

The \trento\ model parameterizes the initial entropy density via the reduced thickness function~\cite{Moreland:2014oya}:
\begin{ceqn}
\begin{equation}
  s = s_0 \left( \frac{\T_A^p + \T_B^p}{2} \right)^{1/p},
  \label{eq:genmean}
\end{equation}
\end{ceqn}
where $\T(x, y)$ is the modified participant thickness function $\T(x, y)
= \sum_{i=1}^{N_\text{part}} \gamma_i\, T_p(x - x_i, y - y_i)$ and $ \gamma_i$ is a random weighting factor. $T_{p}$ is the nucleon thickness function with a Gaussian form:
$T_p(x, y) = \frac{1}{2\pi w^2} \exp(\!-\frac{x^2 + y^2}{2 w^2})$ and $w$ is a tunable effective nucleon width. $s_0$
is a normalization factor and $p$ is a tunable parameter, which makes \trento\ model effectively interpolates
among different entropy deposition schemes, such as {\tt{KLN}}, {\tt{EKRT}}, {\tt WN}, and so on~\cite{Moreland:2014oya,Bernhard:2016tnd}. Following~\cite{Bernhard:2016tnd}, we input a temperature dependent specific shear viscosity $\eta/s(T)$ and specific bulk viscosity $\zeta/s(T)$ for the simulations with \trento\ initial condition.  In Ref.~\cite{Bernhard:2016tnd}, the specific shear viscosity $\eta/s(T)$ above $T_c$ was assumed to be a linear function with tunable minimum value and slope parameter. The specific bulk viscosity $\zeta/s(T)$ was taken a peak form with two functions falls off exponentially at each side, together with a tunable overall normalization factor. Using Bayesian statistics, the free parameters of \trento\ , the initial time $\tau_0$, switching temperature $T_{sw}$, and the parameterized $\eta/s(T)$ and $\zeta/s(T)$ in {\tt{iEBE-VISHNU}} simulations, are simultaneously tuned through the massive data fitting of final multiplicity, mean $p_T$ and integrated flow harmonics $v_n$ in 2.76 A TeV Pb+Pb collisions.  Such massive data evaluation prefer $\tau_0=0.6 \ \mathrm{fm/c}$, $T_{sw}=148 \ \mathrm{MeV}$, together with the extracted $\eta/s(T)$ and $\zeta/s(T)$ curves shown in Fig.~1 (denoted as para-I). Other well calibrated parameters for \trento\ initial condition can be found in table IV in~\cite{Bernhard:2016tnd}.

In this paper, we will study and predict various flow observables in both 2.76 A TeV and 5.02 A TeV Pb+Pb collisions. As shown in Fig.~2, the final multiplicities only increase by $\sim$ 30\% after the collision energy raised from 2.76 A TeV to 5.02 A TeV, which corresponds to $\sim$ 10\% increase of the initial temperature. We thus use the same $\eta/s(T)$ and $\zeta/s(T)$ parametrization as well as other related parameter sets extracted in~\cite{Bernhard:2016tnd}, except for re-tuning the normalization factor $s_0$ in Eq.~(4) to fit the final multiplicities of all charged hadrons in 5.02 A TeV Pb+Pb collisions~\footnote{The centralities here and the ones for following calculations in Sec.~4 are all cut by the distributions of all charged hadrons with $|\eta|<0.5$.}. We found that such parameter set-ups could equally well describe the measured flow harmonics of all charged hadrons in both 2.76 A TeV and 5.02 A TeV Pb+Pb collisions (please refer to Sec.~4 for details).  \\[0.10in]

The {\tt{AMPT}} initial condition\cite{Xu:2016hmp,Bhalerao:2015iya,Pang:2012he} constructs the initial energy density profiles through the energy decompositions of individual partons via a Gaussian smearing:
%
\begin{equation}
	\epsilon = K\sum_{i}\frac{E_{i}^{*}}{2\pi\sigma^{2}\tau_{0}\Delta\eta_{s}}\exp\left.(-\frac{(x-x_{i})^{2}+(y-y_{i})^{2}}{2\sigma^2}\right.), \label{eq:epsilon}
\end{equation}
%
where $\sigma$ is the Gaussian smearing factor, $E_{i}^{*}$ is the Lorentz invariant energy of the produced partons
and $K$ is an additional normalization factor. For simplicity, the initial flow are neglected as Ref.~\cite{Xu:2016hmp,Bhalerao:2015iya,Pang:2012he} and the total produced partons from {\tt{AMPT}} are truncated within $|\eta|<1$ to construct the initial energy density profiles in the transverse plane according to the Eq.(\ref{eq:epsilon}).

Following~\cite{Xu:2016hmp}, we input a constant QGP specific shear viscosity and zero specific bulk viscosity, and set the parameters for the pre-equilibrium {\tt AMPT} evolution as: Lund string fragmentation $a$=2.2 and $b$=0.5, strong coupling constant $\alpha=0.4714$ and the screening mass $\mu$=3.226 fm$^{-1}$. Again, considering that the final multiplicities from 2.76 A TeV to 5.02 A TeV Pb+Pb collisions only increase by $\sim$ 30\%, we use the same hydrodynamic starting time $\tau_0=0.6\mathrm{fm}/c$, transport coefficients $\eta/s=0.08$, $\zeta/s=0$ (denoted as para-II in Fig.~1) and Gaussian smearing factor $\sigma=0.6$, and switching temperature $T_{sw}=148 \ \mathrm{MeV}$, but only tune the normalization factor K of the initial condition to fit the final multiplicities of all charged hadrons in 2.76 A and 5.02 A TeV Pb+Pb collisions. We found such parameter set-ups can nicely fit the  multiplicity, $p_T$-spectra and integrated flow harmonics $v_n$ of all charged hadrons at these two collision energies (please also refer to Sec.~4). The details of parameter tuning can be found in our earlier paper~\cite{Xu:2016hmp}.

\begin{figure}[t]
  \centering\includegraphics[scale=0.38]{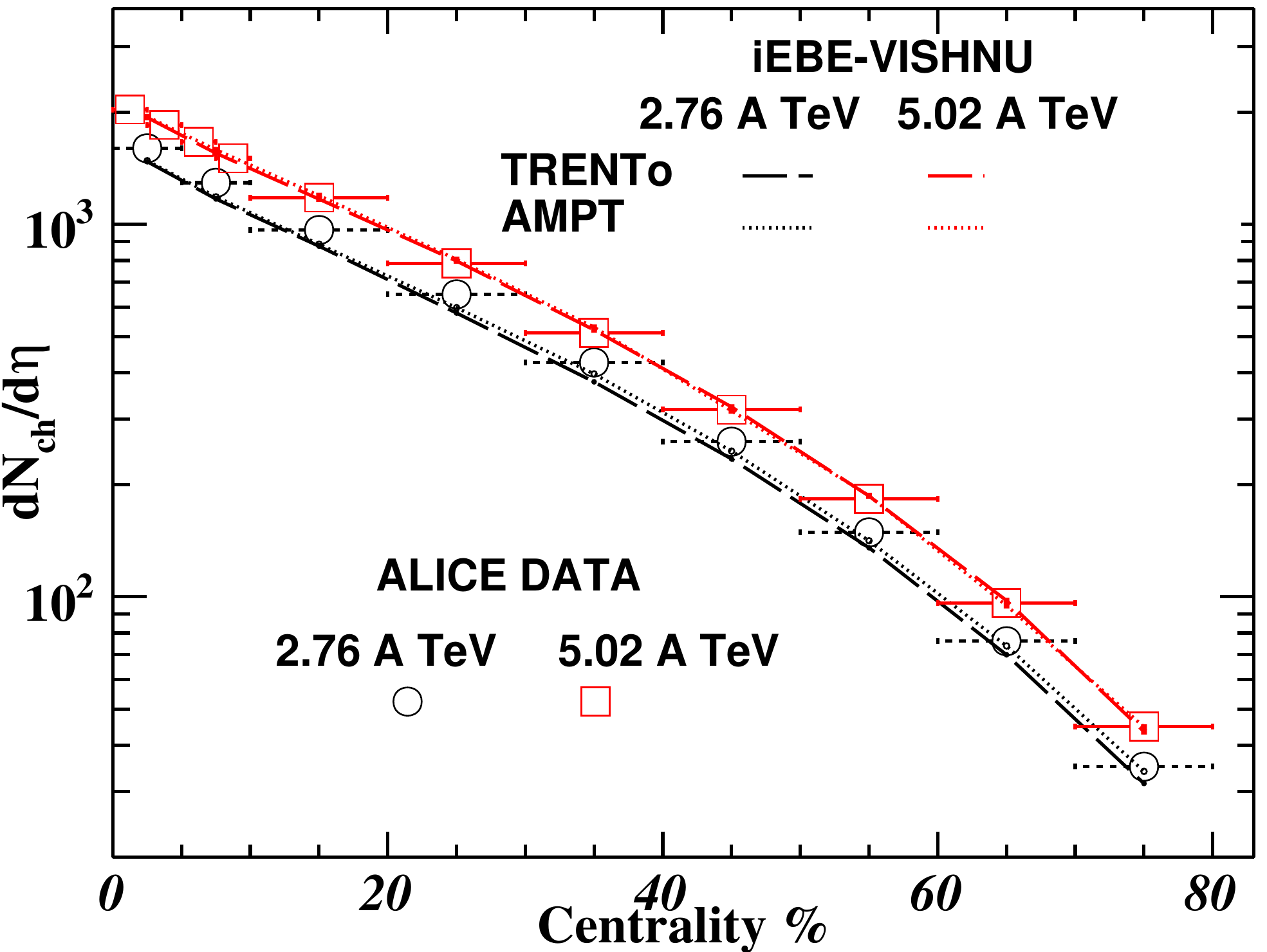}
   \vspace{4mm}
  \caption{(Color online) The centrality dependence of the charged-hadron multiplicity density $dN_{ch}/d\eta$ ($|\eta|<0.5$) in Pb + Pb collisions at 2.76 A TeV and 5.02 A TeV, calculated from {\tt iEBE-VISHNU} with \trento\ and {\tt AMPT} initial conditions. The experimental data are taken from \cite{Aamodt:2010cz} and \cite{Adam:2015ptt}, respectively. }
  \label{dNch}
\end{figure}

\section{Flow observables }

\quad \ In this section, we will briefly introduce the calculation of various flow observables that will be shown in the next section, which include flow harmonics $v_n$,  event-by-event $v_n$ distributions, event-plane correlations,  the Symmetric Cumulants, non-linear response coefficients, and the $p_T$-dependent factorization ratios.

\vspace{0.3cm}
\noindent
\underline{{Flow harmonics and the Q-cumuant method}}
\vspace{0.2cm}

The flow harmonics measure the anisotropy of momentum distributions of final produced hadrons. It can be obtained from a Fourier expansion of the event-averaged azimuthal particle distributions~\cite{Voloshin:1994mz}:
\begin{ceqn}
\begin{eqnarray}
\begin{aligned}
\frac{{\rm d} N}{{\rm d} \varphi}& =\frac{1}{2\pi}  \sum_{-\infty}^{\infty}{V}_n e^{-in\varphi}\\
&= \frac{1}{2\pi} (1+ 2 \sum_{n=1}^{\infty} v_{n} e^{-in(\varphi-\Psi_{n})}),	
\end{aligned}
\end{eqnarray}
\end{ceqn}
where ${V}_n$ is the $n$-th order flow-vector, defined as ${V}_n =v_ne^{in\Psi_n}$,  $v_{n} = \langle cos \, n(\varphi - \Psi_n) \rangle$ is the n-th flow harmonics and $\Psi_{n}$ is the corresponding event plane.

The generally used Q-cumulant method~\cite{Bilandzic:2010jr} measures the flow harmonics $v_n$ from 2- and multi-particle correlations without the knowledge of the event plane. The $Q_n$-vector is defined as:
\begin{ceqn}
\begin{equation}
Q_{n} = \sum_{i=1}^{M}e^{in\varphi_{i}},
\end{equation}
\end{ceqn}
where $M$ is the multiplicity in a single event and $\varphi_i$ is the azimuthal angle of the emitted particle $i$. With this $Q_n$-vector, the 2-and 4-particle azimuthal correlations in a single event can be calculated as~\cite{Bilandzic:2010jr}:
\begin{equation}
\begin{aligned}
\langle 2 \rangle_{n,-n} & = \frac{|Q_{n}|^{2} - M} {M(M-1)}, \\
\langle 4 \rangle_{n,n,-n,-n} & = \frac{|Q_{n}|^{4} + |Q_{2n}|^{2} - 2 \cdot {\rm{Re}}[Q_{2n}Q_{n}^{*}Q_{n}^{*}]  } { M(M-1)(M-2)(M-3) }  \\
& \ \  ~ ~ - 2 \frac{ 2(M-2) \cdot |Q_{n}|^{2} - M(M-3) } { M(M-1)(M-2)(M-3) },
\end{aligned}
\label{Eq:Mean24}
\end{equation}

Here, we have used the general notation of the single-event k-particle correlators $\langle k \rangle_{n_{1}, n_{2}, ..., n_{k}} \equiv \langle \cos(n_1\varphi_1\! + \!n_2\varphi_2\!+\!\cdots\!+\!n_i\varphi_i) \rangle  \,(n_1\geq n_2 \geq \cdots \geq n_i)$ and $\langle ... \rangle$ means an average over all the particles in a single event. After averaging over the whole events within the selected centrality bin, the obtained 2- and 4-particle cumulants are:
\begin{ceqn}
\begin{equation}
\begin{aligned}
c_{n}\{2\} & = \langle \langle 2 \rangle \rangle_{n,-n}, \\
c_{n}\{4\} & = \langle \langle 4 \rangle \rangle_{n,n,-n,-n} - 2 \cdot \langle \langle 2 \rangle \rangle^{2}_{n,-n},
\end{aligned}
\label{Eq:c24}
\end{equation}
Then, the 2- and 4-particle integrated flow harmonics can be calculated as~\cite{Bilandzic:2010jr}:
\begin{eqnarray}
v_n\{2\} &= \sqrt{c_n\{2\}}, \quad  v_n\{4\} &= \sqrt[4]{-c_n\{4\}}.
\end{eqnarray}
\end{ceqn}

In general, the 4-particle correlations in flow harmonics $v_n\{4\}$ could largely suppress
the non-flow effects from jets, resonance decays and etc..  However, they still significantly influence $v_n\{2\}$ obtained from the 2-particle correlations. To suppress such non-flow effects, one divides the whole event into two sub-events with a certain pseudorapidity gap $|\Delta \eta|$, and then calculate the modified 2-particle azimuthal correlations as:
\begin{ceqn}
\begin{equation}
\langle 2 \rangle^{|\Delta \eta|}_{n,-n} = \frac{Q_{n}^{A}Q_{n}^{B *}} {M_{A}M_{B}},
\label{Eq:Mean2Gap}
\end{equation}
\end{ceqn}
where $Q_{n}^{A(B)}$ and $M_{A(B)}$ are the $Q_{n}$-vectors and multiplicities of sub-event A(B). The Q-cumulant and flow harmonics from 2-particle correlations with a $|\Delta \eta|$ gap become:
\begin{ceqn}
\begin{equation}
c_{n}\{2, |\Delta\eta|\} = \langle \langle 2 \rangle \rangle^{|\Delta \eta|}_{n,-n},
 v_n\{2, |\Delta\eta|\} = \sqrt{c_n\{2,|\Delta\eta|\}}.
\label{Eq:v22Gap10}
\end{equation}
\end{ceqn}
One could also define a single-event correlator averaged over the Particles Of Interests (POIs). Such POIs can be some specific identified hadrons or the hadrons within some transverse momentum ranges and so on, depending on the physics interested. With the correlators of POIs, one can further calculate the (differential) flow harmonics flow of all charged hadrons or identified hadrons and etc. in a similar way as described above. Again the non-flow effects can be suppressed by a pseudo rapidity gap $|\Delta \eta|$. For the limited space, we will not further outline the lengthy formulas, but refer to \cite{Bilandzic:2010jr,Zhou:2015iba}
for details.

Note that the Scalar Product (SP) method is also belong to the framework of two-particle
correlations, but uses different event average weights when compared with the
the standard Q-cumulant method~\cite{Bilandzic:2010jr}. We found that, for the {\tt{iEBE-VISHNU}} hybrid model simulations with non-flows mainly contributed from resonance decays, the Q-cumulant method and the scalar product method generate almost identical flow harmonics from semi-central to semi-peripheral collisions~\cite{Xu:2016hmp,Xu:private}

\vspace{0.3cm}
\noindent
\underline{{Distributions of event-by-event flow harmonics}}
\vspace{0.2cm}

The event-by-event $v_n$ distributions reflect the event-by-event fluctuations of the initial states of relativistic heavy-ion collisions, which are not significantly influenced by the hydrodynamic evolution and can provide strong constraints for the initial condition models~\cite{Gale:2012rq,Aad:2013xma,Jia:2013tja}.

In general, one first calculates the per-particle flows from an expansion of the particle distributions in azimuthal angle $\phi$ and then obtains the event-by-event distributions of flow harmonics in a selected centrality bin. However, finite multiplicities and non-flow effects can make the distributions of observed per-particle flow deviate from the true distributions. To suppress such effects, one implements the standard Bayesian unfolding procedure\cite{Aad:2013xma,Adye:2011gm} to obtain the true $v_n$ distributions. For the limited spaces, we do not out-line the details to calculate the $v_n$ distributions and the related Bayesian unfolding procedure, but refer to~\cite{Aad:2013xma,Adye:2011gm} for details.

For a selected centrality bin, the averaged flow harmonics $\langle v_n \rangle$ from model calculations and experimental measurements are not exactly the same,  but exist some differences.  To get rid of such influences and focus on the shape of the $v_n$ distributions, one defines the scaled event-by-event $v_n$ distributions $P(v_n/\langle v_n\rangle)$, which are generally used to evaluate the related model calculations with certain initial conditions~\cite{Gale:2012rq,Aad:2013xma}.

\vspace{0.3cm}
\noindent
\underline{{Event-plane correlations}}
\vspace{0.2cm}

The event-plane correlations evaluate the correlations of various flow angle combinations, which shed lights on the initial state fluctuations and the non-linear response of the evolving system~\cite{Aad:2014fla,Qiu:2012uy,Bhalerao:2013ina,Teaney:2013dta,Bhalerao:2014xra}. Following~\cite{Aad:2014fla,Bhalerao:2013ina}, we implement the Scalar-Product method to calculate the event-plane correlations.  The two and three event-plane correlations are defined as:
\begin{equation}
\begin{split}
& \cos\left[c_1 n_1 \Psi_{n_1} - c_2 n_2 \Psi_{n_2} \right]  \\
 &\qquad = \frac{\langle \tilde{Q}_{n_1A}^{c_1} \tilde{Q}_{n_2B}^{c_2*} \rangle}{\sqrt{\langle \tilde{Q}_{n_1A}^{c_1} \tilde{Q}_{n_1B}^{c_1*}\rangle} \sqrt{\langle \tilde{Q}_{n_2A}^{c_2}\tilde{Q}_{n_2B}^{c_2*}\rangle}} \\
&\cos\left[c_1 n_1 \Psi_{n_1} + c_2 n_2 \Psi_{n_2} - c_3 n_3 \Psi_{n_3} \right]  \\
& \qquad    = \frac{ \langle \tilde{Q}_{n_1A}^{c_1} \tilde{Q}_{n_2A}^{c_2}\tilde{Q}_{n_3B}^{c_3*}\rangle}{\sqrt{\langle \tilde{Q}_{n_1A}^{c_1} \tilde{Q}_{n_1B}^{c_1*} \rangle \langle \tilde{Q}_{n_2A}^{c_2} \tilde{Q}_{n_2B}^{c_2*} \rangle \langle \tilde{Q}_{n_3A}^{c_3}\tilde{Q}_{n_3B}^{c_3*} \rangle}},
\end{split}
\end{equation}

Here, the subscript ``A'' and ``B''  donate the two different sub-events, which are separated by a $|\Delta\eta|$ gap. The reduced flow vector $\tilde{Q}_n$ is defined as:
\begin{ceqn}
\begin{equation}
\label{qcomplex}
\tilde{Q}_n
\equiv\frac{1}{N} \sum_j e^{in\varphi_j},
\end{equation}
\end{ceqn}
where $N$ is the number of particles in a sub-event, and $\varphi_j$ is azimuthal angles of particle $i$. Note that for a specific two or three event-plane correlator, the azimuthal symmetry requires that $c_1n_1-c_2n_2=0$ or $c_1n_1+c_2n_2-c_3n_3=0$~\cite{Aad:2014fla,Bhalerao:2013ina}.


\vspace{0.3cm}
\noindent
\underline{{The Symmetric Cumulant}}
\vspace{0.2cm}

The Symmetric Cumulant $SC(m,n)$ measures the correlations between different flow harmonics, which is defined as ~\cite{ALICE:2016kpq,Bhalerao:2014xra}:
\begin{eqnarray}
SC^v(m,n) &=&\left<\left<\cos(m\varphi_1\!+\!n\varphi_2\!-\!m\varphi_3-\!n\varphi_4)\right>\right>_{c}\nonumber\\
&=& \langle \langle 4 \rangle \rangle_{n,m,-n,-m} -   \langle \langle 2 \rangle \rangle_{n,-n}
\cdot \langle \langle 2 \rangle \rangle_{m,-m}  \nonumber\\
&=&\left<v_{m}^2v_{n}^2\right>-\left<v_{m}^2\right>\left<v_{n}^2\right>.
\label{eq:4p_sc_cumulant}
\end{eqnarray}

Here the symmetric cummulant is only defined with $m\neq n$ with two positive integers $m$ and $n$. The single event 4-particle and 2-particle correlations $\langle 4 \rangle_{n,m,-n,-m}$, $ \langle 2 \rangle_{n,-n}$ and $ \langle 2 \rangle_{m,-n}$ can be expressed in term of the Q-vectors (please refer to~\cite{Bhalerao:2014xra,ALICE:2016kpq} for details), and $\langle \langle ... \rangle \rangle$ denotes an average over all the events.

To evaluate the relative strength of the correlations between different flow harmonics, one defines the Normalized Symmetric Cumulants:
\begin{ceqn}
\begin{eqnarray}
NSC^{v}(m, n) = \frac{SC^{v}(m, n)}{\langle v_{m}^{2}\rangle \langle v_{n}^{2}\rangle},
\end{eqnarray}
\end{ceqn}
where $\langle v_{m}^{2}\rangle$ and $\langle v_{n}^{2}\rangle$ can be calculated by the 2-particle cumulants in Eq.(10). For details, please refer to~\cite{ALICE:2016kpq,Zhu:2016puf}.

\vspace{0.3cm}
\noindent
\underline{{Non-linear response coefficients}}
\vspace{0.2cm}

The non-linear evolution of the QGP fireball leads to the mode-couplings between different flow harmonics, which could be evaluated by the non-linear response coefficients~\cite{Bhalerao:2014xra,Qian:2016fpi,Yan:2015jma}. Except for the second and third order anisotropic flows which are linearly proportional to second and third order eccentricities of the initial state, the higher-order anisotropic flow vectors contain contributions of both linear and nonlinear parts, which can be decomposed
as \cite{Bhalerao:2014xra,Qian:2016fpi,Yan:2015jma}:
\begin{equation}
\begin{split}
  V_4 & = V_{4 L} + \chi_{422} V_2^2, \ \ \
  V_5  = V_{5 L} + \chi_{523} V_2 V_3,\\
  V_6 & = V_{6L} + \chi_{624} V_2 V_{4L} + \chi_{633} V_3^2 + \chi_{6222}  V_2^3\,, \\
  V_7 & =  V_{7L} + \chi_{725} V_2 V_{5L} + \chi_{734} V_3 V_{4L} + \chi_{7223}  V_2^2 V_3.
\end{split}
\label{VlVnl}
\end{equation}

Here, the non-linear terms directly involve the contributions from lower order flow anisotropies and the corresponding coefficients $\chi_{mnl}$ and $\chi_{mnlk}$  are called as the non-linear response coefficients (mode-coupling coefficients). Following~\cite{Yan:2015jma},  we implement the Scalar-Product method to calculate the mode coupling coefficients, which are expressed as:
\begin{ceqn}
\begin{widetext}
\begin{eqnarray}
\chi_{422}   &=  &
      \frac{\bra \tilde{Q}_{4A}
  \tilde{Q}_{2B}^{*}\tilde{Q}_{2B}^{*}\ket}{\bra  \tilde{Q}_{2A}\tilde{Q}_{2A}\tilde{Q}_{2B}^{*}\tilde{Q}_{2B}^{*}  \ket},\quad
\chi_{523} =
      \frac{\bra \tilde{Q}_{5A}
  \tilde{Q}_{2B}^{*}\tilde{Q}_{3B}^{*}\ket}{\bra  \tilde{Q}_{2A}\tilde{Q}_{3A}\tilde{Q}_{2B}^{*}\tilde{Q}_{3B}^{*}  \ket}, \quad
 \chi_{624} =
\frac{ \langle \tilde{Q}_{6A}\tilde{Q}_{2B}^{*}\tilde{Q}_{4B}^{*} \rangle \langle \tilde{Q}_{2A}^{2} \tilde{Q}_{2B}^{*2} \rangle
                                  -  \langle \tilde{Q}_{6A}\tilde{Q}_{2B}^{*3} \rangle \langle \tilde{Q}_{4A}\tilde{Q}_{2B}^{*2} \rangle}
                                   {\bigl( \langle \tilde{Q}_{4A} \tilde{Q}_{4B}^{*} \rangle \langle \tilde{Q}_{2A}^{2} \tilde{Q}_{2B}^{*2} \rangle%
                                  {-}\langle \tilde{Q}_{4A} \tilde{Q}_{2B}^{*2} \rangle^2\bigr)\,\langle \tilde{Q}_{2A} \tilde{Q}_{2B}^{*} \rangle},
\nonumber\\
\chi_{633} &=  & \frac{ \langle \tilde{Q}_{6A}\tilde{Q}_{3B}^{*2} \rangle}{\langle \tilde{Q}_{3A}^{2}\tilde{Q}_{3B}^{*2}) \rangle},\quad
\chi_{6222} = \frac{\langle \tilde{Q}_{6A}\tilde{Q}_{2B}^{*3} \rangle}{\langle (\tilde{Q}_{2A}\tilde{Q}_{2B}^{*})^{3} \rangle},
\quad\mathfrak{}
\chi_{734} =
     \frac{\langle \tilde{Q}_{7A}\tilde{Q}_{3B}^{*}\tilde{Q}_{4B}^{*} \rangle \langle (\tilde{Q}_{2A}\tilde{Q}_{2B}^{*})^2 \rangle
                                   - \langle \tilde{Q}_{7A}\tilde{Q}_{2B}^{*2}\tilde{Q}_{3B}^{*} \rangle \langle \tilde{Q}_{4A}\tilde{Q}_{2B}^{*2} \rangle}
                                   {\bigl(\langle \tilde{Q}_{4A}\tilde{Q}_{4B}^{*} \rangle \langle (\tilde{Q}_{2A}\tilde{Q}_{2B}^{*})^2 \rangle%
                                   {-}\langle \tilde{Q}_{4A}\tilde{Q}_{2B}^{*2}\rangle ^2\bigr)\,\langle \tilde{Q}_{3A}\tilde{Q}_{3B}^{*} \rangle},\quad
\nonumber\\
\chi_{725}& =&\frac{\langle \tilde{Q}_{7A}\tilde{Q}_{2B}^{*}\tilde{Q}_{5B}^{*} \rangle \langle \tilde{Q}_{2A}\tilde{Q}_{2B}^{*}\tilde{Q}_{3A}\tilde{Q}_{3B}^{*} \rangle
                                   - \langle \tilde{Q}_{7A}\tilde{Q}_{2B}^{*2}\tilde{Q}_{3B}^{*} \rangle \langle \tilde{Q}_{5A}\tilde{Q}_{2B}^{*}\tilde{Q}_{3B}^{*} \rangle}
                                   {\bigl(\langle \tilde{Q}_{5A}\tilde{Q}_{5B}^{*} \rangle \langle  \tilde{Q}_{2A}\tilde{Q}_{2B}^{*}\tilde{Q}_{3A}\tilde{Q}_{3B}^{*} \rangle%
                                     {-}\langle \tilde{Q}_{5A}\tilde{Q}_{2B}^{*}\tilde{Q}_{3B}^{*} \rangle^2\bigr)\,\langle \tilde{Q}_{2A}\tilde{Q}_{2B}^{*} \rangle},
\chi_{7223} = \frac{ \langle \tilde{Q}_{7A}\tilde{Q}_{2B}^{*2}\tilde{Q}_{3B}^{*} \rangle}{\langle \tilde{Q}_{2A}^{2}\tilde{Q}_{2B}^{*2}\tilde{Q}_{3A}\tilde{Q}_{3B}^{*} \rangle}.
\label{Q_chi}
\end{eqnarray}
\end{widetext}
\end{ceqn}
Here, the whole event is divided into two sub-events, A and B, with a $|\Delta\eta|$ gap separation to suppress the non-flow effects. The reduced flow vectors $\tilde{Q}_{nA}$ and $\tilde{Q}_{nB}$ are defined by Eq.(14), and
$\langle ... \rangle$ means averaging over the whole events, and then taking the real parts.
\begin{figure*}[tbh]
	\begin{centering}
		\centering\includegraphics[scale=0.91]{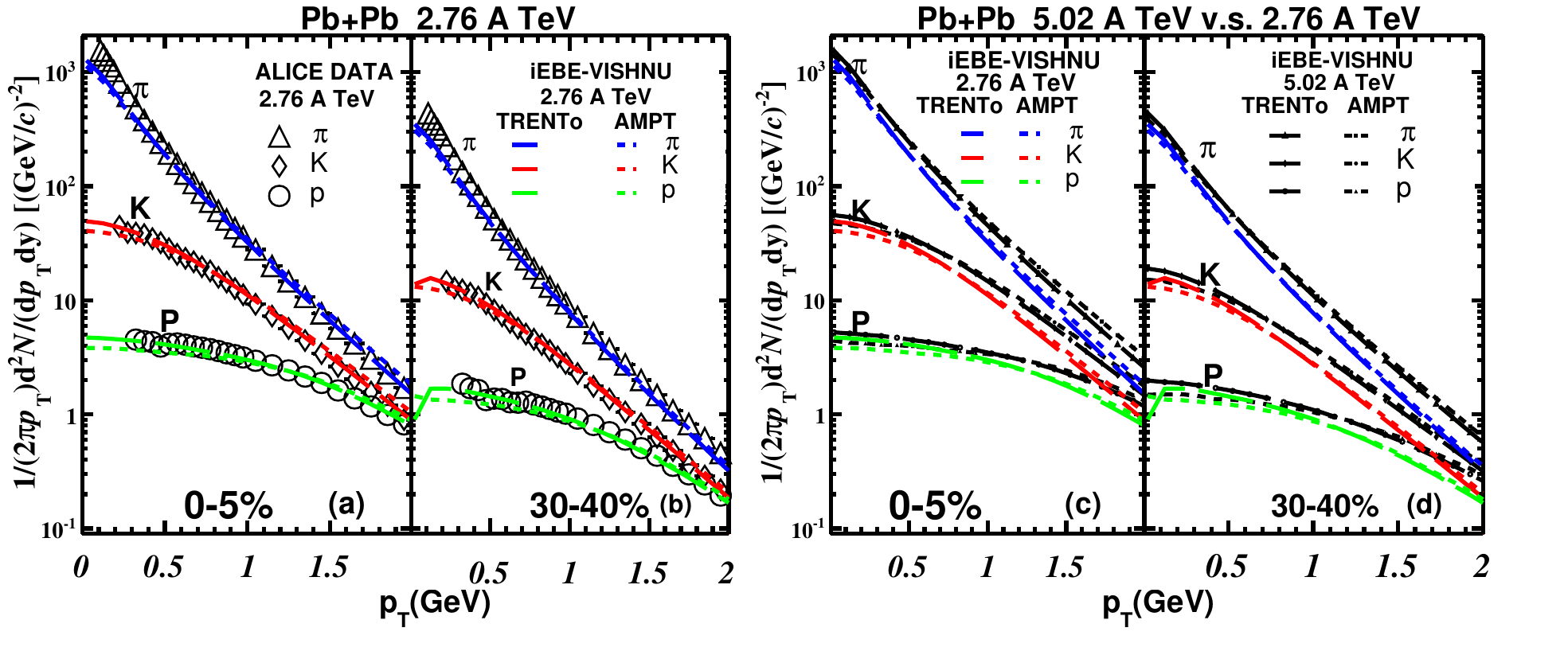}
	\end{centering}
    \vspace{-5mm}
	\caption{(Color online) $p_T$ spectra of pions, kaons, and protons in 0-5\% and 30-40\%  Pb+Pb collisions at 2.76 A TeV and 5.02 A TeV, calculated from {\tt{iEBE-VISHNU}} with  \trento\ and {\tt AMPT} initial conditions.  The experimental data at 2.76 A TeV are taken from {the ALICE paper}~\cite{Abelev:2013vea}.
\label{ptspectra} }
\end{figure*}

\vspace{0.3cm}
\noindent
\underline{{$p_T$-dependent factorization ratio}}
\vspace{0.2cm}

The produced hadrons at different transverse momentum $p_T$ do not share a common flow angle, which leads to the break-up of the flow harmonics factorizations. To evaluate the strength of such break-ups, one defines the $p_T$-dependent factorization ratio~\cite{Heinz:2013bua,Khachatryan:2015oea}:
\begin{ceqn}
\begin{equation}
\label{correlation}
r_n(p_T^a, p_T^b)\equiv\frac{V_{n\Delta}(p_T^a,p_T^b)}{\sqrt{V_{n\Delta}(p_T^a,p_T^a)V_{n\Delta}(p_T^b,p_T^b)}},
\end{equation}
\end{ceqn}

Here, $V_{n\Delta}$ are the average value of $cos(n\Delta\phi)$ for all particles pairs within a momentum bin range, together with a $|\Delta\eta|$ gap to reduce the non-flow effects. It can be calculated as\cite{Khachatryan:2015oea}:
\begin{ceqn}
\begin{equation}
\label{average}
V_{n\Delta} \equiv \avg{\cos(n\dphi)}=\langle\tilde Q^{a(b)}_n \tilde Q^{a(b)*}_n\rangle,
\end{equation}
\end{ceqn}
where $\avg{\ ... }$ denotes averaging over all particle pairs in a single event and then taking an average over all events. $\tilde Q^{a(b)}_n$ is the reduced flow vector of POIs calculated within a specific $p^{a(b)}_T$ bin and
rapidity range: $\tilde{Q}^{a(b)}_n\equiv\frac{1}{N} \sum_j e^{in\varphi_j}$. The related average $\langle ... \rangle$ means averaging over the whole events, and then taking the real parts.

\section{RESULTS AND DISCUSSIONS}

\quad \ Before studying and predicting various flow observables, it is important to check the $p_T$ spectra of identified hadrons since it reflects the radial flow of the expanding system.  Fig.~3 shows the $p_T$ spectra of pions, kaons, and protons in 0-5\% and 30-40\% Pb+Pb collisions at 2.76 A TeV and 5.02 A TeV. The left two panels compare {\tt{iEBE-VISHNU}} calculations with the ALICE data~\cite{Abelev:2013vea} at 2.76 A TeV. For both \trento\ and {\tt AMPT} initial conditions, {\tt{iEBE-VISHNU}} nicely fit the data for these two selected centrality bins, which indicates that hybrid model simulations generate  proper amounts of radial flow. Note that the slope of the $p_T$ spectra is sensitive to the initial time $\tau_0$ and the switching temperature $T_{switch}$. The massive data evaluations from early {iEBE-VISHNU} simulations with \trento\ initial conditions prefer $T_{switch}=148 \ \mathrm{MeV}$ and $\tau_0=0.6 \ \mathrm{fm/c}$ in 2.76 A TeV Pb+Pb collisions~\cite{Bernhard:2016tnd}. For simulations with the {\tt AMPT} initial conditions, we continue to use  the same values of $T_{switch}$ and $\tau_0$. This leads to slightly softer $p_T$ spectra for protons and slightly harder $p_T$ spectra for pions compared with the results obtained with the \trento\ initial conditions, but still make an overall good fit of the measured data below 2 GeV.

Fig.~3 (c) and (d) show the {\tt VISHNU} predictions for the $p_T$-spectra of pions, kaons and protons in 5.02 A TeV Pb + Pb collisions. As introduced in Sec.~II, we use almost the same parameter sets as the ones at 2.76 A TeV, except for tuning the normalization factors of the initial entropy/energy densities to achieve a nice fit of the final multiplicities of all charged hadrons in 5.02 A TeV Pb+Pb collisions. Panels (c) and (d) show that the $p_T$-spectra in 5.02 A TeV are higher and flatter than ones in 2.76 A TeV, which illustrates that stronger radial flow has been developed in the systems with larger final multiplicities at the higher collision energy.
\begin{figure*}[tbh]
	\centering\includegraphics[width=0.8400\linewidth]{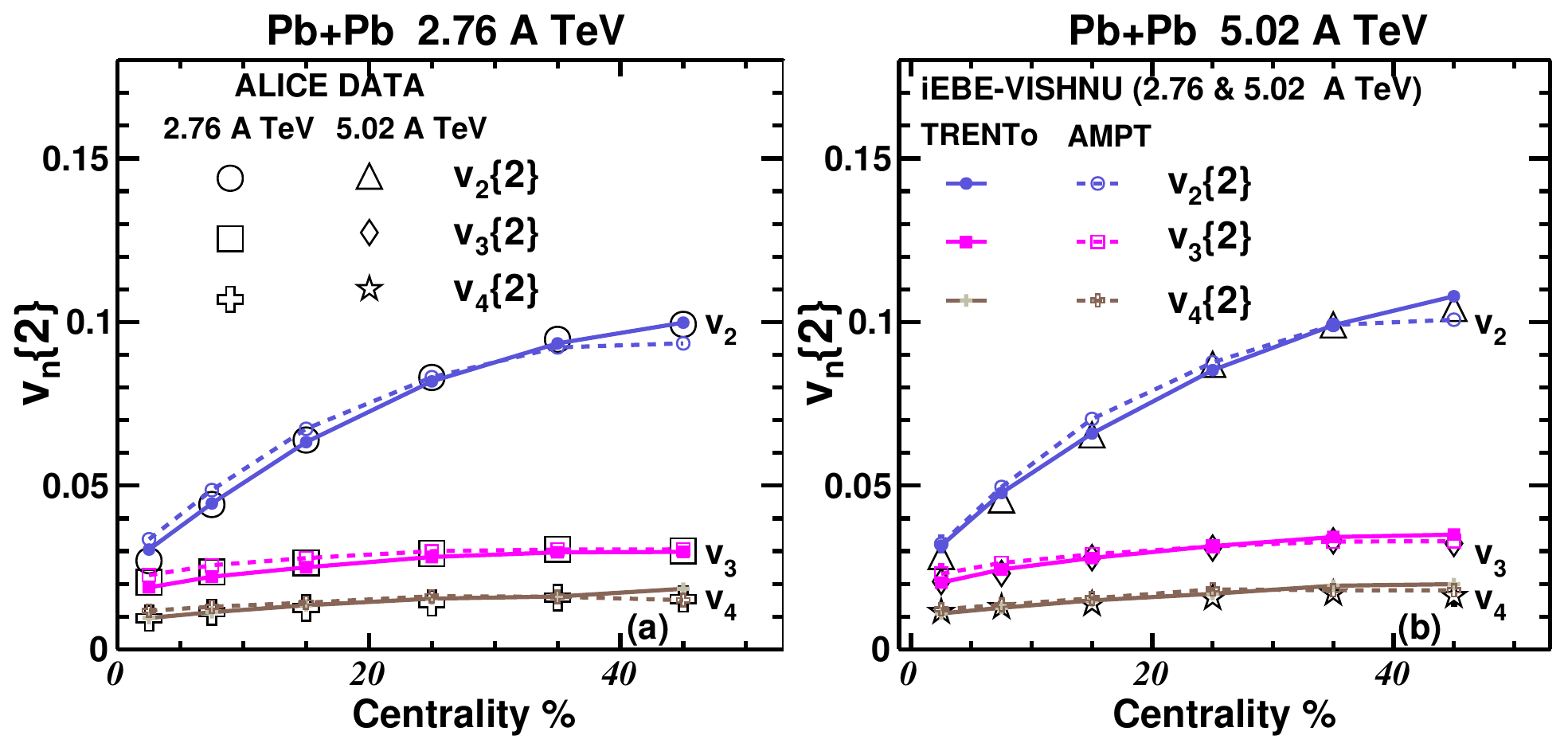}
	\caption{(Color online) Integrated flow $v_n$ (n=2-4) of all charged hadrons in Pb + Pb collisions at 2.76 A TeV (left panel) and 5.02 A TeV (right panel), calculated from {\tt{iEBE-VISHNU}} with  \trento\ and {\tt AMPT} initial conditions.  The experimental data are taken from~\cite{ALICE:2011ab} and~\cite{Adam:2016izf}, respectively.
         \label{integrated_vn}
         }
\end{figure*}

Fig.~4 shows the integrated flow harmonics $v_n$ (n=2-4) of all charged hadrons in 2.76 A TeV and 5.02 A TeV Pb + Pb collisions. Following~\cite{ALICE:2011ab} and~\cite{Adam:2016izf}, we calculate the flow harmonics $v_n$ using the 2-particle cumulant method within $0.2<p_T<5.0\ \mathrm{GeV}$ and $|\eta|<0.8$, together with a pseudo rapidity gap $|\Delta \eta|>1.0$.
For both \trento\ and {\tt AMPT} initial conditions, the transport coefficients and other related parameters in {\tt{iEBE-VISHNU}} have been fine tuned to fit the flow harmonics $v_n$ in 2.76 A TeV Pb+Pb collisions (please refer to Sec.II for details). We found, with the extracted $\eta/s(T)$ and $\zeta/s(T)$ (para-I in Fig.~1) for \trento\ initial condition and $\eta/s=0.08$ and $\zeta/s=0$ (para-II in Fig.~1) for {\tt AMPT} initial condition, {\tt iEBE-VISHNU} can nicely describe the centrality dependent flow harmonics $v_n$ in both 2.76 A TeV and 5.02 A TeV Pb+Pb collisions. The comparison runs in~\cite{McDonald:2016vlt} also showed that, with the same sets of transport coefficients, MUSIC+IP-Glasma simulations can nicely fit the $v_n$ data at these two collision energies. In contrast, the early calculations of the flow harmonics in 200 A GeV Au+Au collisions and 2.76 A TeV Pb+Pb collisions indicated that the average QGP shear viscosity is slightly larger at the LHC than at RHIC, when the final multiplicities increase by about a factor of two~\cite{Song:2011qa,Gale:2013da}. In fact, the final multiplicities between 2.76 A TeV to 5.02 A TeV Pb+Pb collisions only differ by $\sim$30\%, which corresponds to $\sim$10\% change of the initial temperature. We thus do not fine-tuning the transport coefficients for each collision energies, but use the same parameter sets. We find that such choice of parameters can simultaneously fit the individual flow harmonics in both 2.76 A TeV and 5.02 A TeV Pb+Pb collisions.
 \begin{figure*}[tbh]
	\centering\includegraphics[width=0.89\linewidth]{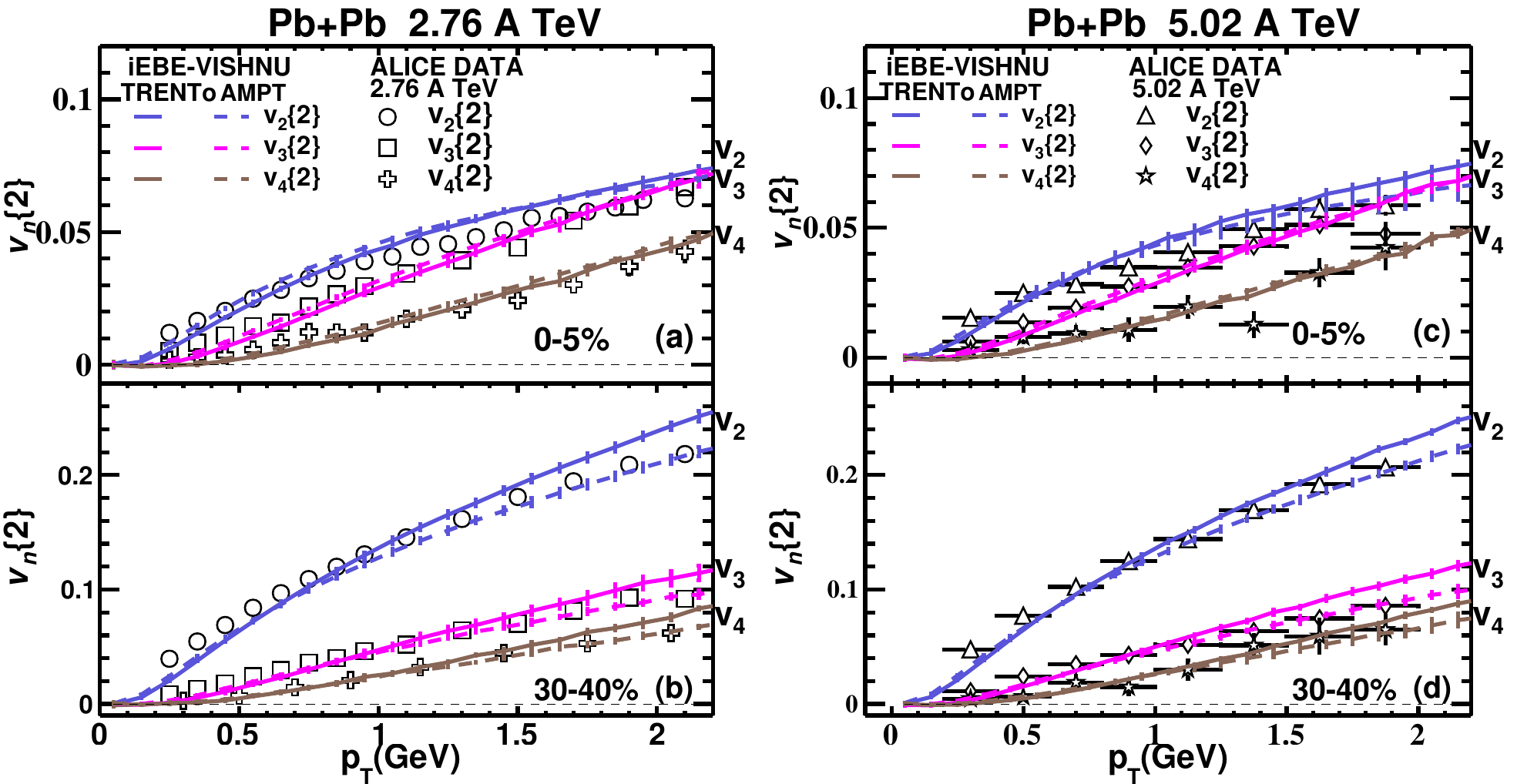}
    \vspace{2mm}
	\caption{(Color online) The differential flow harmonics $v_n(p_T)$ (n=2-4) of all charged hadrons in 0-5\% and 30-40\% Pb + Pb collisions at 2.76 A TeV (left panels) and 5.02 A TeV (right panels), calculated from {\tt{iEBE-VISHNU}} with \trento\ and {\tt AMPT} initial conditions. The experimental data are taken from~\cite{ALICE:2011ab} and~\cite{Adam:2016izf}, respectively.
         \label{vn_total_pt}
         }
\end{figure*}

Fig.~5 shows the differential flow harmonics $v_n(p_T)$ (n=2-4) of all charged hadron in 0-5\% and 30-40\% Pb + Pb collisions at 2.76 A TeV and 5.02 A TeV, calculated by {\tt iEBE-VISHNU} and measured by ALICE using the 2-particle cumulant method
within $|\eta|<0.8$~\footnote{Instead of imposing a pseudorapidity cut $|\Delta\eta|>1.0$ as~\cite{ALICE:2011ab}
and~\cite {Adam:2016izf}, we calculate the 2- particle cumulants using two sub-events with $|\Delta\eta|>0$ in order to reduce the error bars of the limited {\tt{iEBE-VISHNU}} runs. The non-flow effects in {\tt{iEBE-VISHNU}} are dominated by resonance decays. The past simulations~\cite{Xu:2016hmp,XuSong} have shown that the $v_n(p_T)$ curves with $|\Delta\eta|>0$ and $|\Delta\eta|>0.8$ cuts almost overlap.}.  For \trento\ initial conditions, {\tt iEBE-VISHNU} roughly fit the ALICE data in these two collision energies, but with slightly larger slopes. This leads to over-predictions of the $v_n(p_T)$ data above 1 GeV, especially for the 30-40\% centrality. In fact, the parameter sets used in our calculations were obtained from the massive data fitting of the particle yields, mean $p_T$ and integrated flow harmonics in 2.76 A TeV Pb+Pb collisions~\cite{Bernhard:2016tnd}.  Considering the relatively larger error bars, the differential flow harmonics $v_n(p_T)$ were not included in the early massive data evaluations. This partially explains why the current {\tt iEBE-VISHNU} simulations with \trento\ initial conditions do not perfectly describe the $v_n(p_T)$ data. Note that the MUSIC + {\tt IP-Glasma} simulations~\cite{McDonald:2016vlt} also over-predicted the slope of the $v_n(p_T)$ curves and did not very nicely fit the $v_n(p_T)$ data in both 2.76 A TeV and 5.02 A TeV Pb+Pb collisions. Compared with these two simulations, {\tt iEBE-VISHNU} with {\tt{AMPT}} initial condition gives a better description of the data, especially for 30-40\% centrality. We have also noticed that $v_n(p_T)$ data below 0.5 GeV are all slightly under-predicted for these simulations with different initial conditions. In~\cite{Adam:2016nfo}, it was pointed out that the $v_n(p_T)$ data at lower $p_T$ region may contaminated by residual non-flow effects, which have not been fully removed.

Fig.~6 shows the differential flow harmonics $v_n(p_T)$ (n=2-4) of identified hadrons in $10-20\%$ and  $30-40\%$ Pb+Pb collisions at 2.76 A TeV and 5.02 A TeV. Following~\cite{Adam:2016nfo}, we calculate $v_n(p_T)$ using the Scalar Product method with particle of interest (POIs) and reference particles (RPs) selected from two sub-events within $-0.8<\eta<0$ and $0<\eta<0.8$. Note that the ALICE data at 2.76 A TeV~\cite{Adam:2016nfo} have further subtracted the residue non-flow effects using the corrections from p--p collisions. This is not necessary for our {\tt iEBE-VISHNU} calculations since the related non-flow effects are mainly from resonance decays.  The left panels (a-c) compare our model calculations with the data in 2.76 A TeV Pb+Pb collisions. For \trento\ initial condition, {\tt iEBE-VISHNU} can roughly describe the $v_n(p_T)$ of pions, kaons and protons at 10-20\% centrality, but over-predicts the $v_n(p_T)$ above 1 GeV at 30-40\% centrality. For {\tt{AMPT}} initial condition, {\tt{iEBE-VISHNU}} gives an overall quantitative description of the ALICE data for these two selected centrality bins. The situation is similar to the case in Fig.~5 since $v_n(p_T)$ of identified hadrons reflect both the total momentum anisotropies and their distributions among various hadron species.

\begin{figure*}[tbh]
\begin{centering}
	\centering\includegraphics[scale=0.86]{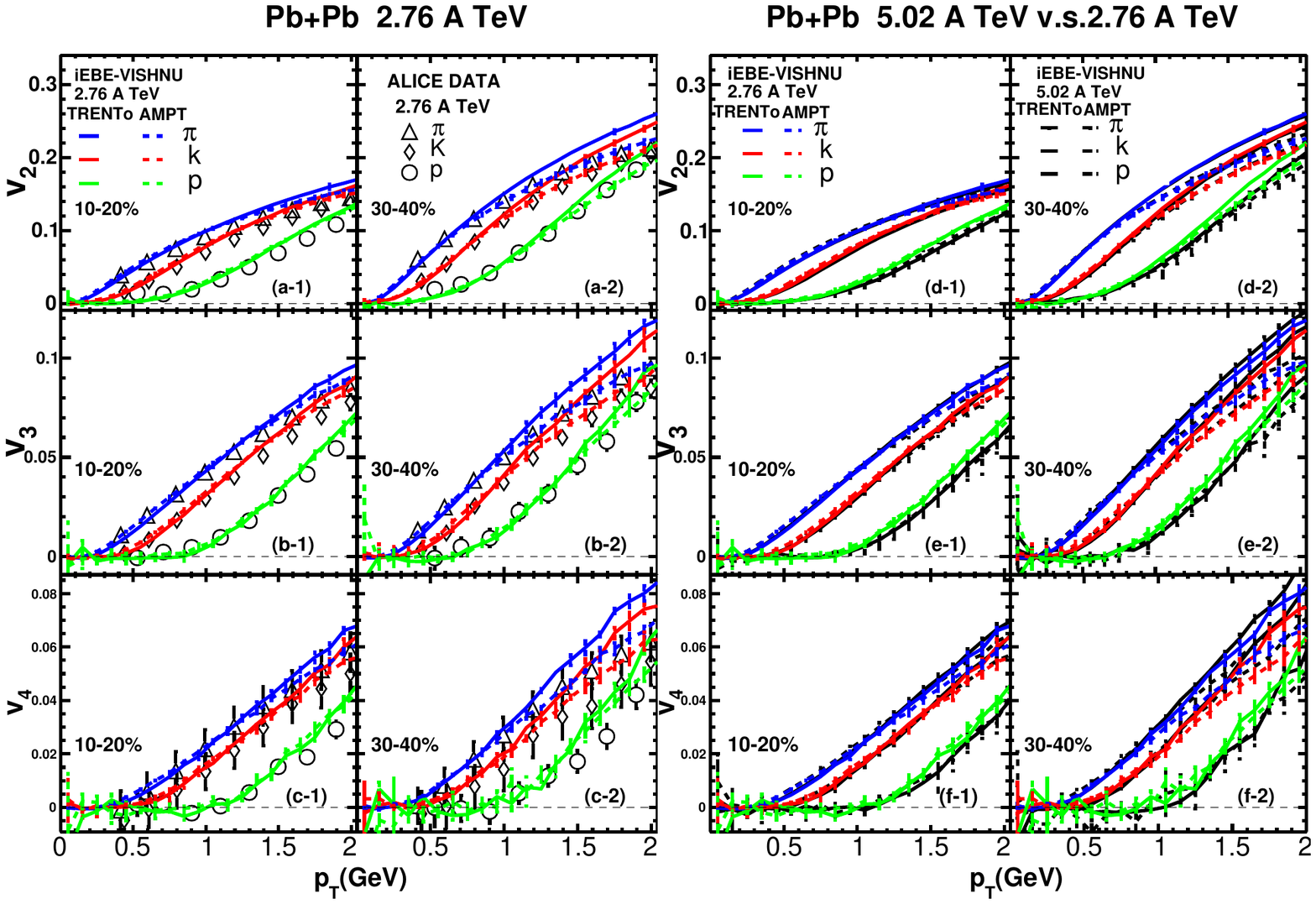}
    \vspace{2mm}
    \end{centering}
    \vspace{-1mm}
\caption{(Color online) The differential flow harmonics $v_n(p_T)$ (n=2-4) of pions, kaons and protons in 10-20\% an 30-40\% Pb+Pb collisions at 2.76 A TeV (left panels) and 5.02 A TeV (right panels), calculated from {\tt{iEBE-VISHNU}} with  \trento\ and {\tt AMPT} initial conditions. The experimental data at 2.76 A TeV are taken from~\cite{Adam:2016nfo}.
	\label{vn_pt_mix} }
\end{figure*}
\begin{figure*}[tbh]
    \begin{centering}
	\includegraphics[width=0.990\linewidth]{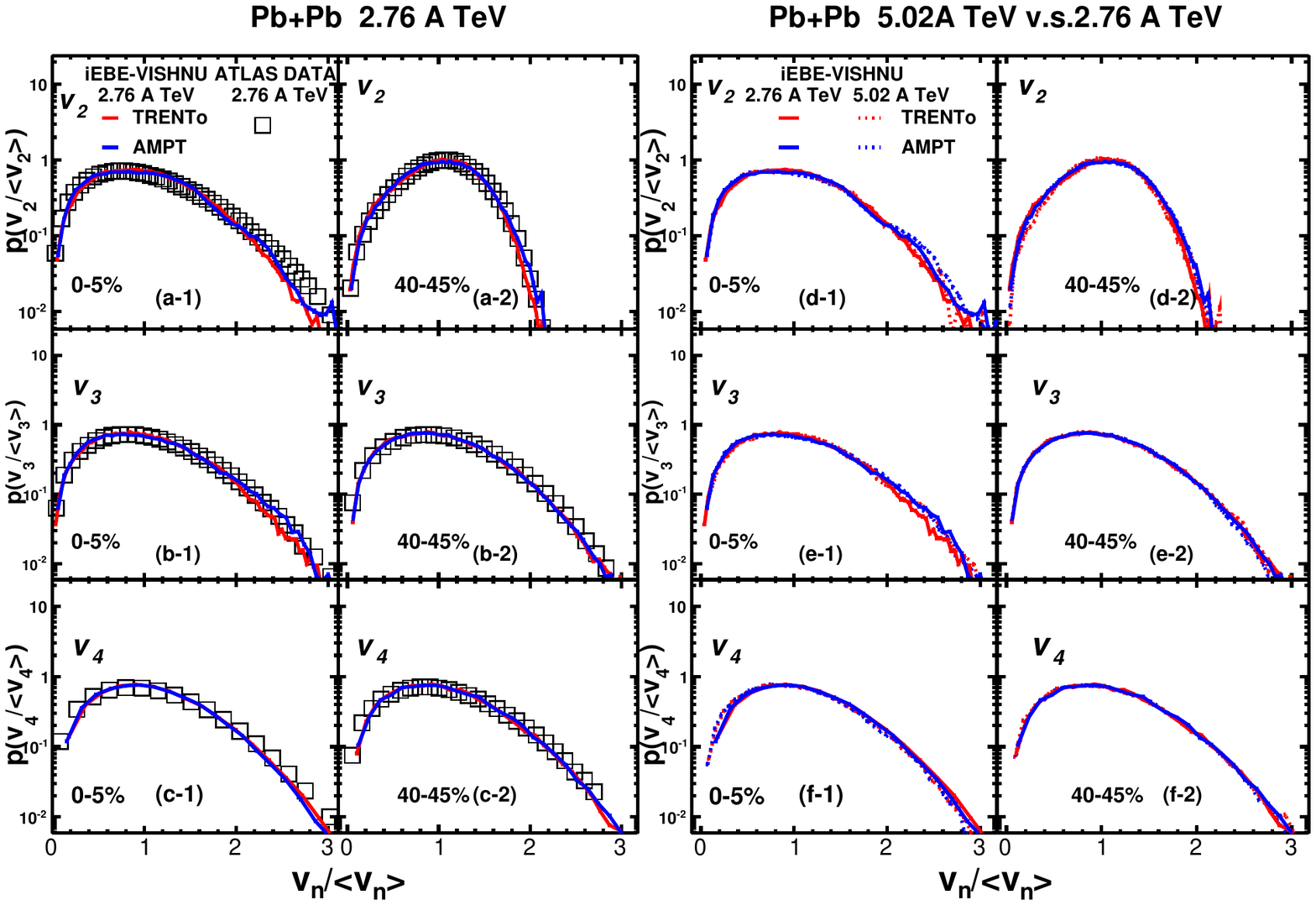}
    \end{centering}
	\caption{(Color online) The scaled event-by-event  $v_n$ distributions in  $0-5\%$  and $40-45\%$   Pb + Pb  collisions at 2.76 A TeV and 5.02 A TeV, calculated from {\tt{iEBE-VISHNU}} with  \trento\ and {\tt AMPT} initial conditions.  The experimental data at 2.76 A TeV are taken from the ATLAS paper~\cite{Aad:2013xma}.
         \label{vn_dis}
         }
\end{figure*}
 \begin{figure*}[tbh]
     \begin{centering}
	\includegraphics[width=1.0\linewidth]{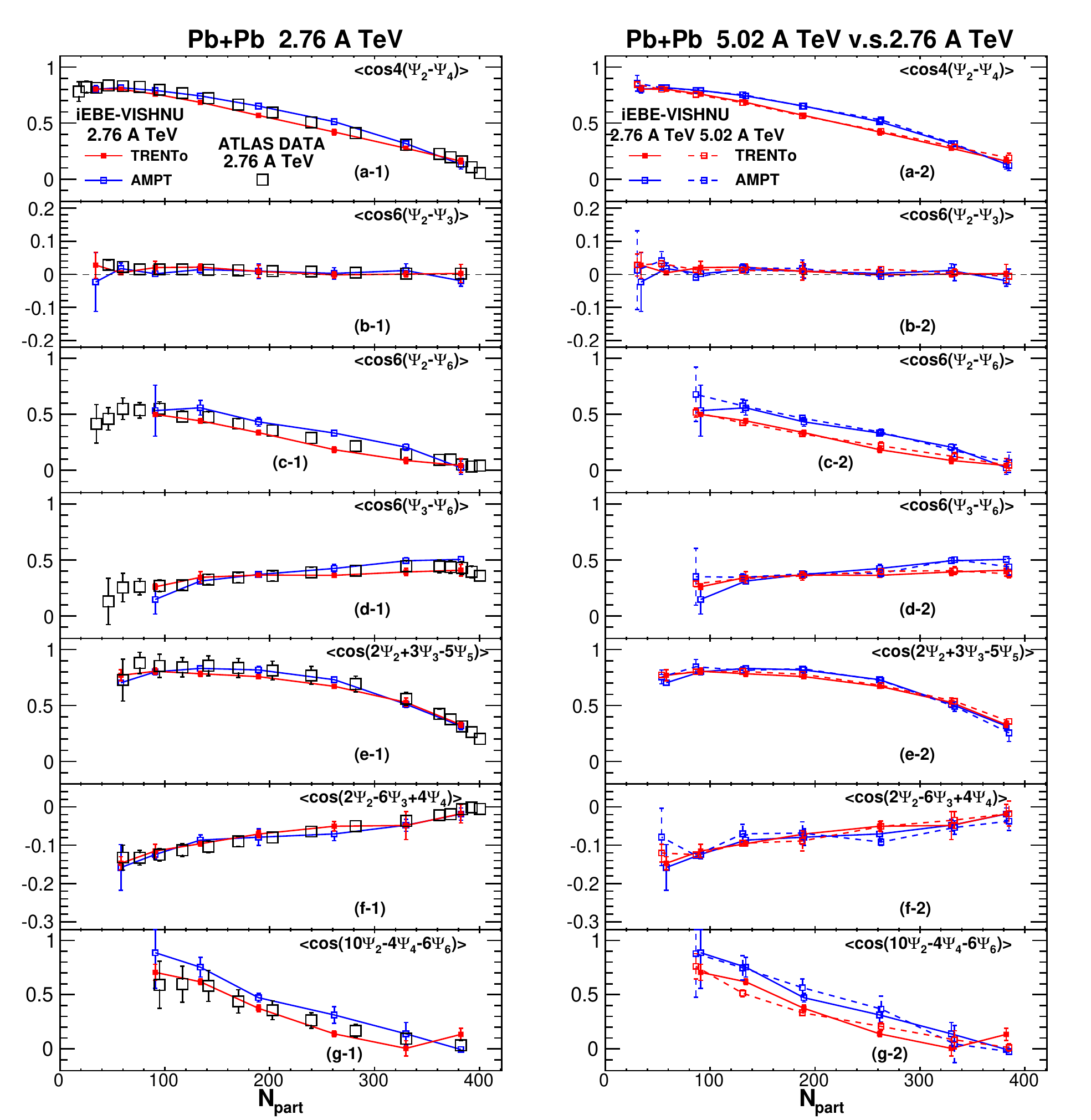}
     \end{centering}
	\caption{(Color online)  Event-plane correlations as a function of participant number in Pb+ Pb collisions at 2.76 A TeV and 5.02 A TeV, calculated from {\tt{iEBE-VISHNU}} with  \trento\ and {\tt AMPT} initial conditions. The data at 2.76 A TeV are taken from {the ATLAS paper}~\cite{Aad:2014fla}, and the $N_{part}$ values are taken from~\cite{Aad:2014fla,Adam:2015ptt}.
         \label{flow_angle}
         }
\end{figure*}

In the right panels (d-f), we predict  $v_n(p_T)$ (n=2-4) of pions, kaons and protons in 5.02 A TeV Pb + Pb collisions, together with a comparison to the {\tt iEBE-VISHNU} results in 2.76 A TeV Pb + Pb collisions. For both \trento\  and {\tt AMPT} initial conditions, the differences between these two collision energies are pretty small, which also show similar $v_n$ mass-orderings.  Note that the measured and calculated $v_n(p_T)$ (n=2-4) of all charged hadrons also almost overlap between these two collision energies (please refer to Fig.~\ref{vn_total_pt} in this paper and Fig.~2 in~\cite{Adam:2016izf}). The early comparison of the flow harmonics at RHIC and the LHC has shown that $v_2(p_T)$ of all charged hadrons almost overlap, while the $v_2$ mass splittings between pions and protons are enlarged with the increase of collision energy\cite{Snellings:2014vqa}. As shown in Fig.~6, the $v_n$ mass-splittings between pions and protons slightly increase from 2.76 A TeV to 5.02 A TeV due to the slightly increased of radial flow.

In Ref.~\cite{McDonald:2016vlt}, the differential flow harmonics $v_2(p_T)$ of $\Lambda$, $\Xi$ and $\phi$ have also been predicted,  which presented certain mass-ordering patterns among these strange and multi-strange hadrons. While, other early research showed that the $v_2$ mass-orderings between $\Lambda$ and p are largely influenced by the pre-equilibrium flow~\cite{Heinz:2015arc} and the magnitude of the $v_2^\phi$ is sensitive to the interaction between the $\phi$ meson and the hadronic matter~\cite{Song:2013qma}. Considering these complexities and the requirement of much higher statistical runs for the model calculations, we do not further predict $v_n(p_T)$ of these strange and multi-strange hadrons, but leave it to future study.

Fig.~7 shows the scaled event-by-event $v_n$ distributions (n=2-4)  in 0-5\% and 40-45\% Pb + Pb  collisions at 2.76 A TeV
and 5.02 A TeV.  Following~\cite{Aad:2013xma}, we first calculate the integrated $v_2$
within transverse momentum $p_T>0.5$ GeV and pseudorapidity $|\eta|<2.5$, using the single-particle method, and then perform the standard Bayesian unfolding procedure~\cite{Adye:2011gm,Aad:2013xma} to obtain the ``true'' $v_n$ distributions.
The left panels (a-c) compare the measured and calculated scaled $v_n$ distributions $P(v_n/\langle v_n\rangle)$ in 2.76 A TeV Pb+Pb collisions. For both {\tt{AMPT}} and \trento\ initial conditions, {\tt iEBE-VISHNU} nicely describes the measured $P(v_n/\langle v_n\rangle)$ curves from ATLAS. As observed in~\cite{Gale:2012rq},  the scaled $v_n$ distributions follow the the scaled $\varepsilon_n$ distributions, for n=2 and 3, due to the linear hydrodynamic response. For n=4, the scaled $\varepsilon_n$ distributions show small deviations from the experimental data in semi-central Pb+Pb collisions~\cite{Gale:2012rq}. The non-linear hydrodynamic evolution coupling the modes between n=2 and n=4, leading to a nice description of the $P(v_n/\langle v_n\rangle)$ data for n=4.

The right panels (d-f) show {\tt iEBE-VISNU} predictions for the scaled $v_n$ distributions in $0-5\%$ and $40-45\%$ Pb + Pb collisions at 5.02 A TeV, together with a comparison with the results at 2.76 A TeV. For both \trento\ and {\tt AMPT} initial conditions, the $P(v_n/\langle v_n\rangle)$ curves at these two collisions energies overlap with each other. As discussed in the above text, the scaled $v_n$ distributions mostly follow the scaled $\varepsilon_n$ distributions, which thus are  insensitive to the collision energy. \\

\par Fig.~8 shows the event-plane correlations as a function of participants number in Pb+ Pb collisions at 2.76 A TeV and 5.02  A TeV. Following the ATLAS paper~\cite{Aad:2014fla}, we calculate the event-plane correlations using the scalar product method
with a pseudorapidity  gap $|\Delta\eta|>1.0$ and
within $p_T>$0.5 GeV and $|\eta|<2.5$. The left panels show that, for both \trento\ and {\tt AMPT} initial conditions, {\tt{iEBE-VISHNU}} can roughly reproduce the ATLAS data in 2.76 A TeV Pb + Pb collision~\footnote{For the limited space, we do not plot the whole 14 event-plane correlations as measured in experiments, but only show 7 representative correlations.}. More specifically, our model calculations nicely describe the decreasing trends of $\langle cos4(\Psi_2-\Psi_4)\rangle$, $\langle cos6(\Psi_2-\Psi_6)\rangle$, $\langle cos(2\Psi_2+3\Psi_3-5\Psi_5)\rangle$ and $\langle cos(10\Psi_2-4\Psi_4-6\Psi_6)\rangle$,  and the increasing trends of  $\langle cos6(\Psi_3-\Psi_6)\rangle$ and $\langle cos(2\Psi_2-6\Psi_3+4\Psi_4)\rangle$ with the increase of the participant number, which also shows close to zero values for $\langle cos6(\Psi_2$-$\Psi_3)\rangle$,  as measured in experiments.  In Ref.~\cite{Qiu:2012uy}, it was found that the non-linear mode couplings and the related event-plane rotations during the hydrodynamic evolution are essential for a qualitative description of various centrality-dependent correlations, which even flip the signs of some correlators between initial and final states. Their calculations also showed that event-plane correlations are sensitive to both initial conditions and the QGP shear viscosity~\cite{Qiu:2012uy}. However the early {\tt VISH2+1} calculations, with either {\tt{MC-Glauber}} or {\tt{MC-KLN}} initial conditions, failed to quantitatively describe all the measured event-plane correlation data.  In fact, both of these two initial conditions also have difficulties to fit all the flow harmonics $v_n$ as well as the event-by-event $v_n$ distributions~\cite{Aad:2013xma,Shen:2014lye}. Compared with the early investigations, our {\tt{iEBE-VISHNU}} simulations with \trento\ and {\tt AMPT} initial conditions could nicely describe the data of individual flow harmonics, which also largely improve the description of the event-plane correlations.  Similarly, the recent {\tt MUSIC} simulations with the successful {\tt IP-Glasma} initial condition, also nicely described these measured event-plane correlations~\cite{McDonald:2016vlt}.

The right panels of Fig.~8 show the {\tt iEBE-VISHNU} predictions on the event-plane correlations in 5.02 A TeV Pb +Pb collisions, which almost overlap with the corresponding ones at 2.76 A TeV. Some of the correlators $\langle cos4(\Psi_2-\Psi_4)\rangle$, $\langle cos6(\Psi_2-\Psi_6)\rangle$, $\langle cos(10\Psi_2-4\Psi_4-6\Psi_6)\rangle$ and etc. shows certain separations for \trento\ and {\tt AMPT} initial conditions, but insensitive to the collision energy. This indicates that the hydrodynamic responses of the corresponding initial correlations are similar at these two collision energies.

Fig.~9 shows the Symmetric Cumulants $SC^v(4,2)$ and $SC^v(3,2)$
and Normalized Symmetric Cumulants $NSC^v(4,2)$ and $NSC^v(3,2)$ in Pb + Pb collisions at 2.76 A TeV and 5.02 A TeV~\footnote{Other symmetric Cumulants $SC^v(4,3)$ and $SC^v(5,2)$ $SC^v(5,2)$ can also be predicted, using the same {\tt iEBE-VISHNU} simulations. However, the related Normalized Symmetric Cumulants $NSC^v(4,3)$ and $NSC^v(5,2)$ $NSC^v(5,2)$ require much higher statistical runs to reduce the error bars. Therefore, we do not further predict them here.  For related investigations, please refer to~\cite{Zhu:2016puf}.}. Following \cite{ALICE:2016kpq}, these Symmetric Cumulants are calculated by the Q-cumulant method within $0.2<p_T<5.0 \ \mathrm{GeV}$ and $|\eta|<0.8$. The left panels compare our model calculations with the experimental data in 2.76 A TeV Pb + Pb collisions. For both \trento\ and {\tt AMPT} initial conditions, {\tt{iEBE-VISHNU}} could roughly describe the centrality dependent $SC^v(m,n)$ and $NSC^v(m,n)$, which also indicate that $v_2$ and $v_4$ are correlated and $v_2$ and $v_3$ are anti-correlated. In Ref.~\cite{Gardim:2016nrr}, it was pointed out that both centrality bin width and non-trivial event weighting influence the measured and calculated Symmetric Cumulants.  A quantitative description of the SC(m,n) and NSC(m,n) data should further consider these factors, which we would like to leave them to future study.

\begin{figure*}
   \begin{centering}
	\includegraphics[width=0.95\linewidth]{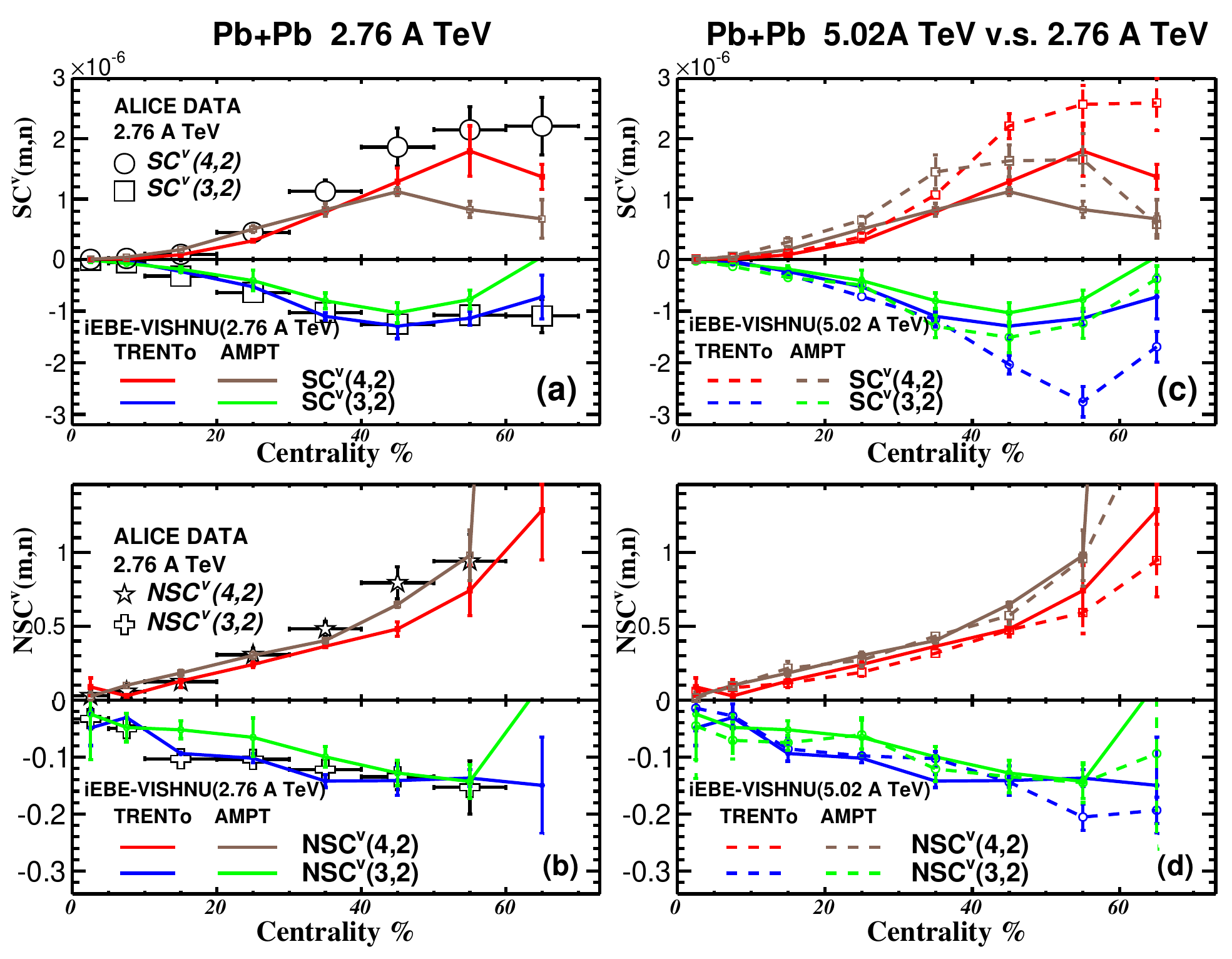}
    \end{centering}
	\caption{(Color online) Symmetric Cumulants $SC^v(m,n)$ and Normalized Symmetric Cumulants $NSC^v(m,n)$ in 2.76 A TeV and 5.02 A TeV Pb + Pb collisions, calculated from {\tt{iEBE-VISHNU}} with  \trento\ and {\tt AMPT} initial conditions. The $SC^v(3,2)$ and $SC^v(4,2)$ data in 2.76 A TeV Pb+Pb collisions are taken from {the ALICE paper}~\cite{ALICE:2016kpq}.
         \label{vn_vn_276}
         }
\end{figure*}

The right  panels of Fig.~9 show the {\tt iEBE-VISHNU} predictions for the Symmetric Cumulants $SC^v(4,2)$ and $SC^v(3,2)$ and the Normalized Symmetric Cumulants  $NSC^v(4,2)$ and $NSC^v(3,2)$ in 5.02 A TeV Pb + Pb collisions. Due to the slightly larger integrated flow harmonics, the absolute values of $SC^v(4,2)$ and $SC^v(3,2)$ also increase from 2.76 A TeV to 5.02 A TeV Pb+Pb collisions, while the Normalized Symmetric Cumulant $NSC^v(4,2)$ and $NSC^v(3,2)$ do not significantly change with the collision energy. In~\cite{Zhu:2016puf},  it was pointed out that the  $NSC^v(3,2)$ is mainly determined by the $NSC^\varepsilon(3,2)$ from the initial state due to the linear response $v_2\propto\varepsilon_2$ and $v_3\propto\varepsilon_3$. Due to the mode coupling between $v_2$ and $v_4$, $NSC^v(4,2)$ is influenced by both initial condition and the non-linear evolution of the systems.  Here we find that $NSC^v(4,2)$ shows certain sensitivity to the
initial conditions, but do not significantly change with the collision energy even the hydrodynamic evolution time increases.


In Fig.~10, we predict the centrality dependent non-linear response coefficients in  Pb + Pb collisions at 2.76 A TeV and 5.02 A TeV, using  {\tt{iEBE-VISHNU}} hybrid model with \trento\  and {\tt AMPT}  initial conditions. These non-linear response coefficients are calculated according to the scalar-product formula Eq. (\ref{Q_chi}) with two sub-events divided by a pseudorapiduty gap $|\Delta\eta|>0.8$ and within 0.3$<p_T<$3.0 GeV and $|\eta|<2.4$.  For the collision energies at both 2.76 A TeV and 5.02 A TeV, these non-linear response coefficients present weak centrality dependence, except for the $\chi_{7223}$.
As found in the early paper~\cite{Qian:2016fpi}, these non-linear response coefficients exhibit certain sensitivity to the initial condition.  For example,  $\chi_{523}$, $\chi_{624}$ and $\chi_{723}$  show clear separations for \trento\  and {\tt AMPT}  initial conditions. On the other hand, the non-linear response coefficients, except for  $\chi_{7223}$, are not sensitive to these two collision energies in our model calculations.

\begin{figure*} [tbh]
	\begin{centering}
		\includegraphics[scale=0.86]{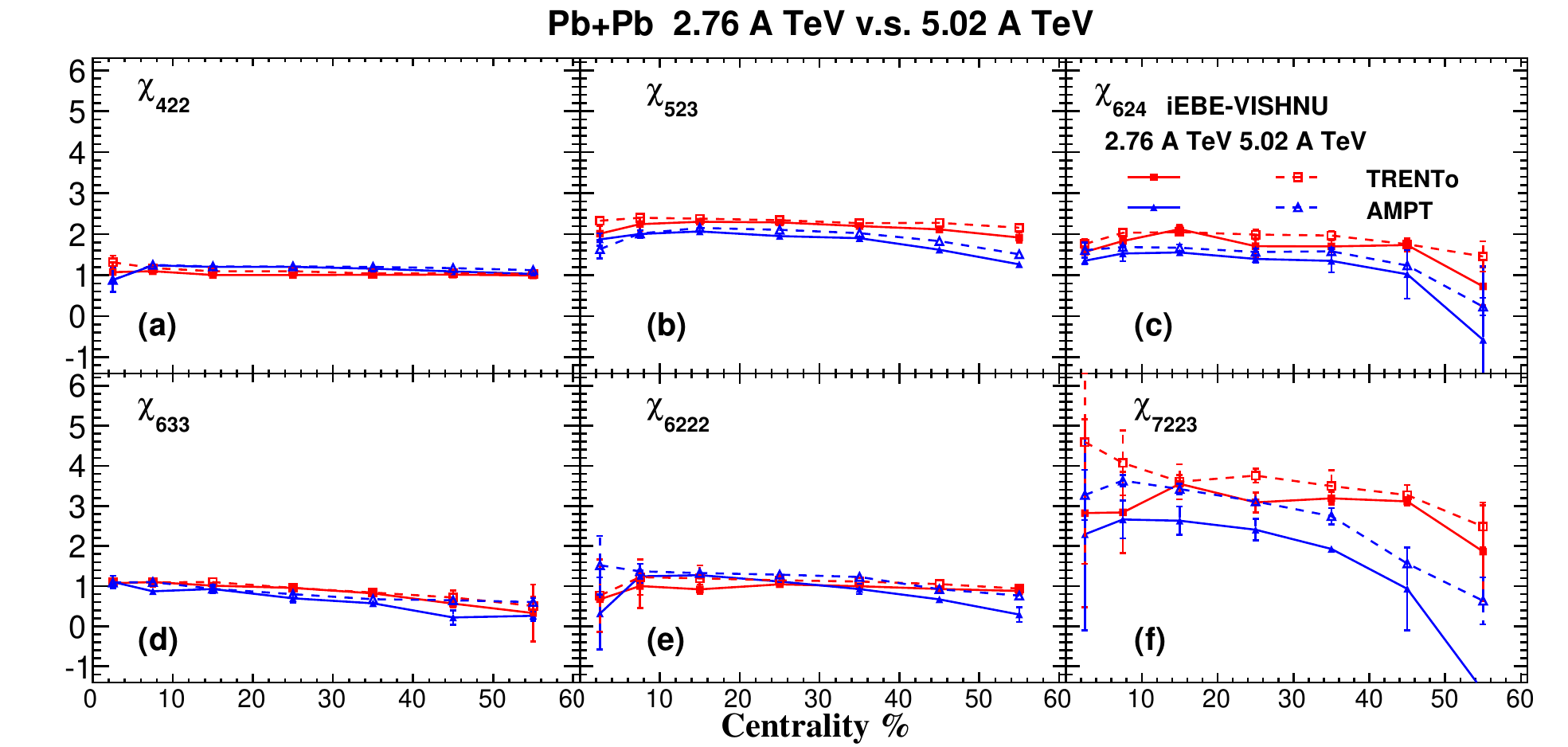}
	\end{centering}
    \vspace{-1mm}
	\caption{(Color online) {\tt{iEBE-VISHNU}} predictions for the centrality dependence of the non-linear response coefficients in 2.76 A TeV and 5.02 A TeV Pb + Pb collisions.
	\label{chi_results} }
\end{figure*}

Fig.~11 shows the $p_T$-dependent factorization ratios, $r_2$ and $r_3$, as a function of $p_T^a-p_T^b$ in $0-5\%$ and $30-40\%$  Pb + Pb collisions at 2.76 A TeV and 5.02 A TeV.  Following~\cite{Khachatryan:2015oea}, we calculate the $p_T$-factorization ratio, $r_2$ and $r_3$, using the scalar-product method with $|\eta^{a,b}|<2.4$ and $|\Delta\eta|>2$. In upper panels, we compare the {\tt{iEBE-VISHNU}} results with the CMS data in 2.76 A TeV Pb+Pb collisions. For both \trento\ and {\tt AMPT} initial conditions, {\tt{iEBE-VISHNU}} hybrid model roughly describe the measured $r_2(p_T^a, p_T^b)$ data in four bins of $p_T^a$.  However,  $r_3(p_T^a, p_T^b)$ from {\tt{iEBE-VISHNU}} drops sharply at larger $p_T^a - p_T^b$ values, which obviously deviates from the CMS data. In~\cite{McDonald:2016vlt}, it was pointed out that
the hadronic rescatterings during the late evolution randomize the flow angles of $v_3$, leading to larger factorization breakings there.

The lower panels show the {\tt iEBE-VISHNU} predictions of $r_2(p_T^a, p_T^b)$ and $r_3(p_T^a, p_T^b)$ in 5.02 A TeV Pb+Pb collisions. We found, for both \trento\ and {\tt AMPT} initial conditions, the values of $r_2$ and $r_3$ are pretty close for the two collision energies at 2.76 A TeV and 5.02 A TeV, which indicate that the non-linear response patterns do not significantly change with the collision energy.
\begin{figure*}
\begin{centering}
		\includegraphics[scale=0.90]{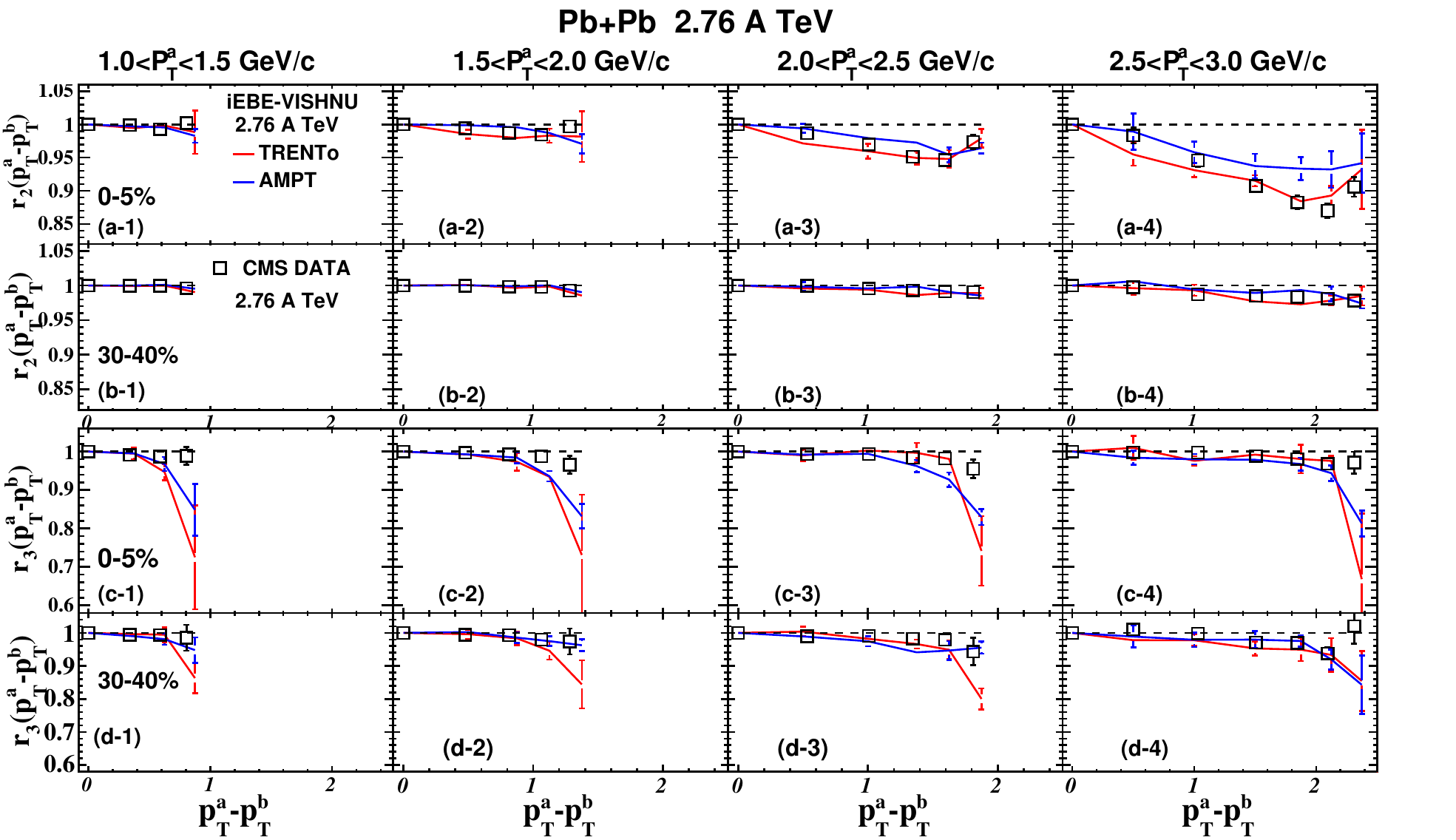}
	\end{centering}
\end{figure*}

\begin{figure*}
\begin{centering}
		\includegraphics[scale=0.900]{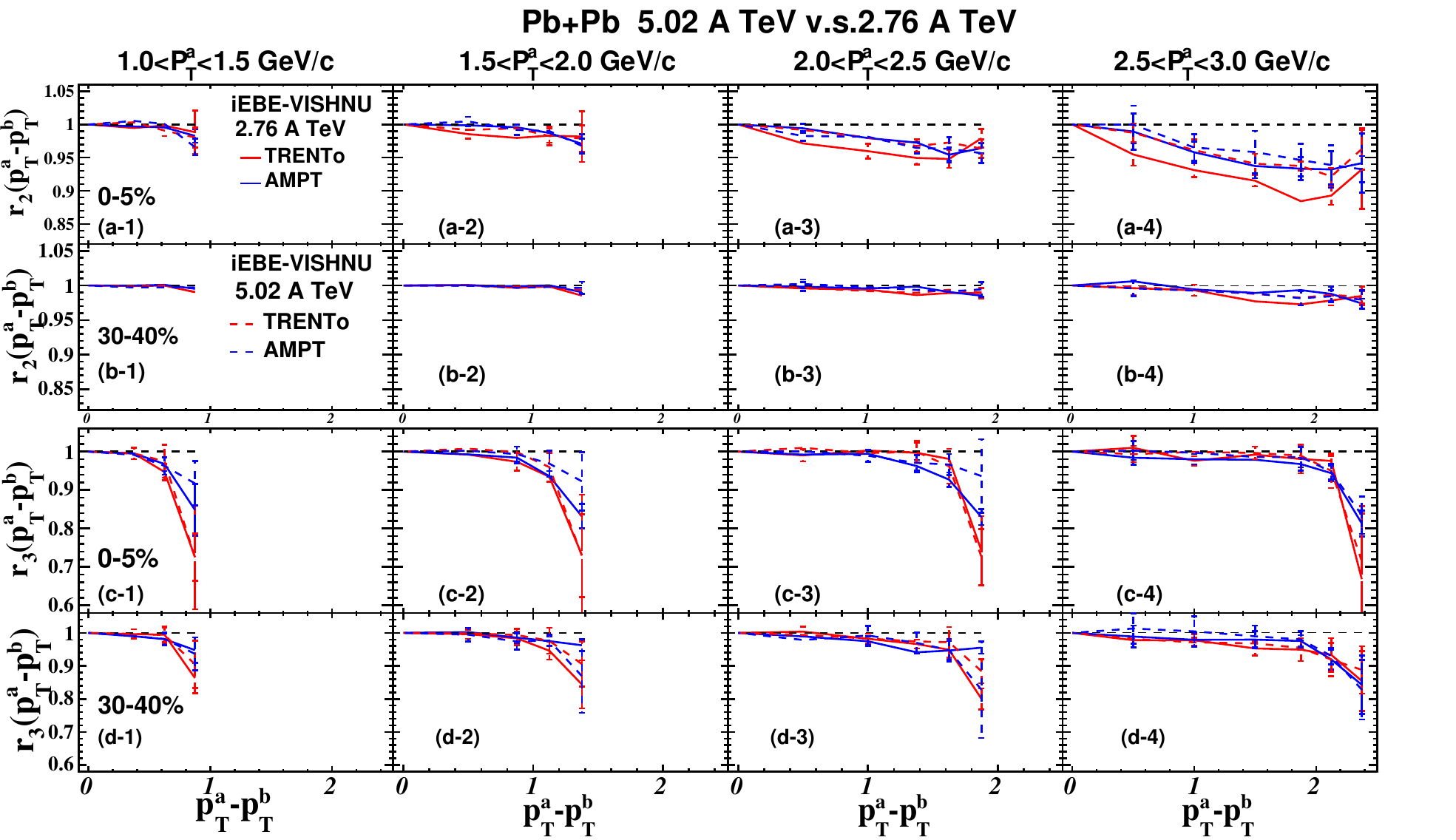}
	\end{centering}
    \vspace{-3mm}
	\caption{(Color online) The factorization ratio, $r_2$ and $r_3$, as a function of $p_T^a-p_T^b$ in $0-5\%$ and $30-40\%$ Pb + Pb collisions at 2.76 A TeV and 5.02 A TeV, calculated from {\tt{iEBE-VISHNU}} with  \trento\ and {\tt AMPT} initial conditions. The experimental data at 2.76 A TeV are taken from the CMS paper~\cite{Khachatryan:2015oea}.
	\label{r_276} }
\end{figure*}

\section{Summary}
\quad \ In this paper,  we studied and predicted various flow observables in Pb +Pb collisions at 2.76 A TeV and 5.02 A TeV, using the {\tt{iEBE-VISHNU}} hybrid model with \trento\ and {\tt AMPT} initial conditions and with different forms of the QGP transport coefficients.   More specifically, we have calculated the integrated and differential flow harmonics
of all charged and identified hadrons, the event-by-event $v_n$ distributions, the event-plane correlations, the correlations between different flow harmonics, the nonlinear response coefficients of higher-order flow harmonics, and $p_T$-dependent factorization ratios. A comparison with the flow measurements in  2.76 A TeV Pb +Pb collisions showed that many of these flow observables can be well described by our model calculations with these two chosen initial conditions, as long as the transport coefficients and other related parameters are properly turned. Some of the flow observables, such as the event plane correlations  $\langle cos4(\Psi_2-\Psi_4)\rangle$ and $\langle cos6(\Psi_2-\Psi_6)\rangle$, the non-linear response coefficients $\chi_{624}$ and $\chi_{723}$, and so on show certain separations for the results obtained with \trento\ and {\tt AMPT} initial conditions. A detailed study of these related flow observables in the future may reveal more details of the initial state fluctuation patterns and the non-linear evolution of the systems.

With almost the same parameter sets, except for the re-tuned normalization factors of initial entropy/energy densities, we predicted various flow observables in 5.02 A TeV Pb+Pb collisions. For the flow harmonics $v_n$ of all charged hadrons, our {\tt iEBE-VISHNU} simulations describe the measured data with the same transport coefficients sets. This indicates that
raising the collision energy from 2.76 A TeV to 5.02 A TeV with the final multiplicities increased by
$\sim$ 30\%, the transport properties of the QGP fireball do not significantly change.  We also predict other flow observables, including $v_n(p_T)$ of identified particles, event-by-event $v_n$ distributions, event-plane correlations, (Normalized) Symmetric Cumulants, non-linear response coefficients and $p_T$-dependent factorization ratios, for 5.02 A TeV Pb+Pb collisions.  We found many of these observables remain approximately the same values as the ones in 2.76 A TeV Pb+Pb collisions. Our theoretical investigations and predictions could shed light to the experimental measurements in the near future.\\

\noindent\textbf{Acknowledgments}:

We thanks the discussion from  A. Behera, J. E. Bernhard, J. Jia, Z. Lin, C. Shen and Y. Zhou. This work is supported by the NSFC and the MOST under grant Nos.11435001, 11675004 and 2015CB856900. H.X. is partially supported by the China Postdoctoral Science Foundation under Grant No. 2015M580908. We gratefully acknowledge the extensive computing resources provided by Super-computing Center of Chinese Academy of Science (SCCAS) and Tianhe-1A from the National Supercomputing Center in Tianjin, China.

\bibliography{V6TeVflow}

\end{document}